\documentclass[12pt,preprint]{aastex62}
\usepackage{verbatim}
\newcommand{\bdv}[1]{\mbox{\boldmath$#1$}}

\def\au{{\rm AU}}

\def\masyr{{\rm mas}\,{\rm yr}^{-1}}
\def\kpc{{\rm kpc}}
\def\mas{{\rm mas}}

\def\muas{\mu{\rm as}}

\def\max{{\rm max}}

\def\rel{{\rm rel}}

\def\eff{{\rm eff}}

\def\e{{\rm E}}
\def\bpi{{\bdv\pi}}
\def\bmu{{\bdv\mu}}

\usepackage{xcolor}
\newcommand{\HL}[1]{\textcolor{black}{#1}}

\begin{document}
\title{KMT-2019-BLG-2073: Fourth Free-Floating-Planet Candidate with $\theta_\e< 10\,\muas$}

\author{\textsc{
Hyoun-Woo Kim$^{1,2}$, 
Kyu-Ha Hwang$^{1}$, 
Andrew Gould$^{3,4}$, 
Jennifer C. Yee$^{5}$, 
Yoon-Hyun Ryu$^{1}$, 
Michael D. Albrow$^{6}$, 
Sun-Ju Chung$^{1,7}$,
Cheongho Han$^{8}$,
Youn Kil Jung$^{1}$,
Chung-Uk Lee$^{1}$, 
In-Gu Shin$^{1}$,
Yossi Shvartzvald$^{9}$,
Weicheng Zang$^{10}$,
Sang-Mok Cha$^{1,11}$, 
Dong-Jin Kim$^{1}$,
Seung-Lee Kim$^{1,7}$, 
Dong-Joo Lee$^{1}$,
Yongseok Lee$^{1,11}$, 
Byeong-Gon Park$^{1,7}$, 
Richard W. Pogge$^{4}$ } }

%----------------------------------------------------------------
\affil{$^{1}$Korea Astronomy and Space Science Institute, 
Daejeon 34055, Republic of Korea}

\affil{$^{2}$Department of Astronomy and Space science, Chungbuk National University, 
Cheongju, 28644, Republic of Korea}

\affil{$^{3}$Max-Planck-Institute for Astronomy, K\"{o}nigstuhl 17, 
D-69117 Heidelberg, Germany}

\affil{$^{4}$Department of Astronomy, Ohio State University, 
140 W. 18th Ave., Columbus, OH 43210, USA}

\affil{$^{5}$Center for Astrophysics $\mid$ Harvard \& Smithsonian, 60 Garden St., Cambridge, MA 02138, USA}

\affil{$^{6}$University of Canterbury, Department of Physics and Astronomy, 
Private Bag 4800, Christchurch 8020, New Zealand}

\affil{$^{7}$University of Science and Technology, 217 Gajeong-ro, Yuseong-gu, 
Daejeon 34113, Republic of Korea}

\affil{$^{8}$Department of Physics, Chungbuk National University, 
Cheongju 28644, Republic of Korea}

\affil{$^{9}$Department of Particle Physics and Astrophysics, 
Weizmann Institute of Science, Rehovot 76100, Israel}

\affil{$^{10}$Department of Astronomy and Tsinghua Centre for Astrophysics, 
Tsinghua University, Beijing 100084, People's Republic of China}

\affil{$^{11}$School of Space Research, Kyung Hee University, 
Yongin, Gyeonggi 17104, Republic of Korea}

\begin{abstract}

We analyze the very short Einstein timescale ($t_\e\simeq 7\,{\rm hr}$)
event KMT-2019-BLG-2073.  Making use of the pronounced finite-source
effects generated by the clump-giant source, we measure the 
Einstein radius $\theta_\e \simeq 4.8\,\muas$, and so infer a mass
%aaa: $M = 59\,M_\oplus (\pi_\rel/0.016 \,\muas)^{-1}$ where -> $M = 59\,M_\oplus (\pi_\rel/16 \,\muas)^{-1}$, where
$M = 59\,M_\oplus (\pi_\rel/16 \,\muas)^{-1}$, where $\pi_\rel$ is the lens-source
relative parallax.  We find no significant evidence for a host of this
%bbb: at sufficiently wide separation... -> at a sufficiently wide separation... 
planetary mass object, though one could be present at sufficiently wide 
separation.  If so, it would be detectable after about 10 years.
This is the fourth isolated microlens with a measured Einstein radius
$\theta_\e<10\,\muas$, which we argue is a useful threshold for a
``likely free-floating planet (FFP)'' candidate.
%aaa: for ``likely free-floating planet (FFP)'' candidate. -> for a ``likely free-floating planet (FFP)'' candidate.
We outline a new approach to constructing a homogeneous sample 
of giant-star finite-source/point-lens (FSPL) events, within which
the subsample of FFP candidates can be statistically analyzed.
We illustrate 
this approach using 2019 KMTNet data and show that there appears to be a large
$\theta_\e$ gap between the two FFP candidates and the 11 other FSPL
events. 
 We argue that such sharp features are more identifiable in
%bbb: a sample selected on... -> a sample selected based on...
a sample selected on $\theta_\e$ compared to the traditional
approach of identifying candidates based on short $t_\e$.

\end{abstract}


\section{{Introduction}
\label{sec:intro}}

Characterizing the free-floating planet (FFP) population is the first step in understanding their origins. FFP candidates were originally identified
from their short Einstein timescales,
\begin{equation}
t_\e \equiv {\theta_\e\over\mu_\rel};
\qquad
\theta_\e\equiv\sqrt{\kappa M\pi_\rel};
\qquad
\kappa\equiv {4 G\over c^2\au}\simeq 8.1\,{\mas\over M_\odot},
\label{eqn:thetae}
\end{equation}
where $\theta_\e$ is the Einstein radius, $M$ is the mass of the lens, 
and $(\pi_\rel,\bmu_\rel)$ are the lens-source relative (parallax, proper motion).
\citet{sumi11} found evidence for a population of low-mass objects
from a bump in the timescale distribution at $t_\e\sim 1\,$day, corresponding
%bbb: analysis of two years of data -> analysis of the two year data
to roughly Jupiter mass objects, from their analysis of two years of data
from the MOA-II survey.  A subsequent analysis of substantially more
data from the OGLE-IV survey by \citet{mroz17} concluded that there was
no such excess of $t_\e\sim 1\,$day events, but they did find an excess of 
shorter, $t_\e\sim 0.2\,$day, events.  According to the $t_\e\propto M^{1/2}$
scaling of Equation~(\ref{eqn:thetae}), this would correspond to roughly
Neptune-mass planets.

However, it is difficult to make inferences about the mass of any
particular microlens based on the Einstein timescale alone, because
this quantity depends on $\pi_\rel$ and $\mu_\rel$ as well as the
lens mass.  That is, at fixed $t_\e$, the mass scales as 
$M\propto \mu_\rel^2/\pi_\rel$ as a function of these unknown quantities.
In particular, $\mu_\rel$ can easily vary by a factor $\sim 5$ for
typical lenses, making inferences about individual objects very uncertain.
Moreover, in contrast to bound planets with luminous hosts, there is
no hope of measuring $\mu_\rel$ from subsequent high resolution observations
for FFPs.
Hence, one must adopt a statistical approach to derive conclusions about
a putative population that generates excess events as a function
of timescale.

A substantial step forward was taken by \citet{ob161540,ob121323,ob190551}
when they measured the angular Einstein radii of four short $t_\e$
events, including three with $\theta_\e< 10\,\muas$, namely
OGLE-2012-BLG-1323, OGLE-2019-BLG-0551, and OGLE-2016-BLG-1540,
with corresponding Einstein radii $\theta_\e=(2.37,4.35,9.2)\,\muas$.
These measurements were made thanks to the fact that, in each case,
the lens transited the source, giving rise to ``finite source effects'',
i.e., deviations from a standard \citet{pac86} light curve.  These
$\theta_\e$ measurements partially break the three-way degeneracy in
Equation~(\ref{eqn:thetae}) by removing $\mu_\rel$ as an unknown.
Hence, for these three lenses, the mass can be estimated
\begin{equation}
M = {\theta_\e^2\over\kappa\pi_\rel}\rightarrow
(14,48,217)M_\oplus\biggl({\pi_\rel\over 16\,\muas}\biggr)^{-1},
\label{eqn:meval}
\end{equation}
where we have scaled to a typical value of $\pi_\rel$ for lenses in the
Galactic bulge.  If the lenses lay in the Galactic disk, the mass
estimates would be lower.

The scaling in Equation~(\ref{eqn:meval}) illustrates that
``$\theta_\e < 10\,\muas$'' is a qualitative indicator of
``good FFP candidate''.  That is, at this boundary, a lens
would have to have $\pi_\rel<1\,\muas$ in order to be in the formal
brown-dwarf regime, $M>13\,M_{\rm jup}$.  Of course, such events 
%aaa: ...can happen, they... -> ...can happen, but they...
(with lens-source distances $D_{LS}\la 10\,$pc) can happen, but they
are extremely rare.  Thus, the appearance of three FSPL events in the
regime $\theta_\e < 10\,\muas$ is strong evidence of a population
of FFPs (or wide-orbit planets for which the host does not
give rise to any signal in the event).

All three of the sources are giant stars, with angular source
radii $\theta_* =(11.9,19.5,15.8)\,\muas$.  If the sources lie in
the bulge at $D_s\sim 8\,\kpc$ (as they almost certainly do based
on their kinematics and the relative probability of their being lensed),
then these correspond to physical source radii $R_* \sim (20,33,27)\, R_\odot$.
This is not accidental.  Accurate measurement of finite-source effects
requires many points over the source crossing time, $t_*=\theta_*/\mu_\rel$,
and the chance of obtaining many measurements is enhanced when $\theta_*$
is large.  In fact, the sources in these three events are exceptionally
large, even for giants.

Here, we report on the discovery of a fourth FFP candidate that satisfies
%aaa: that satisfies the $\theta_\e < 10\,\muas$, KMT-2019-BLG-2073, -> that satisfies the $\theta_\e < 10\,\muas$ criterion, KMT-2019-BLG-2073, 
the criterion
$\theta_\e < 10\,\muas$, KMT-2019-BLG-2073, with $\theta_\e =4.8\,\muas$.
Like the previous three candidates, the lens magnifies a giant star
%aaa: $\theta_*=\,5.4\\muas$ -> $\theta_*=\,5.4\,\muas$
in the Galactic bulge, in this case with $\theta_*=\,5.4\,\muas$.

% ob121323 0.155 +- 0.005 d 2.37 +- 0.10 uas  5.6 +- 0.3  11.9 +- 0.5|same 
% ob170560 0.905 +- 0.005 d 38.7 +- 1.6 uas  15.6 +- 0.7  34.9 +- 1.5|paper
% ob161540 0.320 +- 0.003 d  9.2 +- 0.5 uas  10.5 +- 0.6  15.8 +- 0.8
% ob190551 0.381 +- 0.017 d 4.35 +- 0.34 uas  4.17 +-0.35 19.5 +- 1.6

This discovery prompts us to map out a strategy and carry out the preliminary work for a statistical
study of FFP candidates that have $\theta_\e$ measurements. Such a study, based on $\theta_\e$, has three major advantages over previous efforts using the $t_\e$ distribution to characterize the FFP population.

First, as mentioned above it is subject to significantly less ambiguity in the mass of candidates. For a study based on $t_\e$, the underlying mass distribution will still be convolved with the $\mu_{\rm rel}$ distribution. By contrast, one based on $\theta_\e$ is independent of the $\mu_{\rm rel}$ distribution.

Second, a distribution based on $\theta_\e$ will have an intrinsically higher proportion of FFP candidates compared to one based on $t_\e$.
As pointed out by \citet{gould12} in another context,
the intrinsic rate of point-lens events that display finite-source effects
by any class of objects scales directly with the number density of
this class of objects (assuming that all classes of lenses have the same 
kinematic and physical distribution).  That is, the cross section in the
rate equation for all events is $2\theta_\e$ (which scales 
$\theta_\e\propto M^{1/2}$), while the cross section in the rate equation
for finite-source events is $2\theta_*$, which has no mass dependence.
Thus, low-mass objects, planets
in particular, are favored relative to the general rate
by a factor $(M_{\rm planet}/M_{\rm star})^{-1/2} = 100
[(M_{\rm planet}/M_{\rm Neptune})/(M_{\rm star}/0.5\, M_\odot)]^{-1/2}$.

Third, the relevant timescale of a search for FSPL events is relatively independent of the lens mass, or more precisely, of the direct observable, $\theta_\e$. The detection and characterization of an event requires that the relevant timescale be longer than the interval between observations, with the characterization improving with increasing timescale. For FFP events, $t_\e$ becomes comparable to the cadence of microlensing surveys. So, for a $t_\e$-based search, it becomes increasingly difficult to detect events as the mass (and so $t_\e$) decreases.
By contrast, an FSPL-based search relies only on $t_*$, which is independent of mass and is the lower-limit on the half-width duration of an FSPL event.
 For example, the Einstein timescale of OGLE-2016-BLG-1928 was only $t_\e \approx 0.0288\ \mathrm{days}$ whereas the event itself lasted for an order magnitude longer due to finite source effects \citep{ob190551}. Hence, to a certain limit (see Section \ref{sec:stats}), it is no more difficult to detect and characterize small $\theta_\e$ events than larger ones. As a result, the selection function for a $\theta_\e$-based study of the FFP population is much simpler than for a $t_\e$-based study.

Finally, the fact that the known FSPL FFP candidates all have giant sources, leads us to consider a $\theta_\e$ search restricted to such sources. This empirical observation results, in part, from the fact that the rate at which finite source effects occur is proportional to $\theta_*$. Hence, one advantage of focusing on giant sources is that they have a higher proportion of FSPL events relative to smaller sources. However, restricting the search to giant sources also has a number of other practical implications, which we discuss in more detail in Section \ref{sec:stats}. But first, we turn our attention to KMT-2019-BLG-2073, the inspiration for this search.

\section{{Observations}
\label{sec:obs}}

KMT-2019-BLG-2073, at (RA, Decl.)$_{\rm J2000}$ = (17:49:53.08, $-29$:35:17.30)
[$(l,b) = (-0.07,-1.13)$]
%ccc: Korean... -> Korea...
was announced by the Korea Microlensing Telescope Network 
(KMTNet, \citealt{kmtnet}) AlertFinder system \citep{alertfinder}
as ``clear microlensing'' at UT 04:41 on 14 August 2019 
(HJD$^\prime$ = HJD - 2450000 = 8709.70).  KMTNet is comprised of
three identical 1.6m telescopes with $(2^\circ\times 2^\circ)$ cameras
%ccc: Siding Springs... -> Siding Spring... (from the official website)
%ccc: Cerro Tololo Interamerican... -> Cerro Tololo Inter-American... (from the official website)
at three sites, Siding Spring Observatory, Australia (KMTA),
Cerro Tololo Inter-American Observatory, Chile (KMTC), and
South African Astronomical Observatory, South Africa (KMTS).
The event lies in the overlap region of KMT fields BLG02 and BLG42,
with a combined $I$-band cadence of $\Gamma=4\,{\rm hr}^{-1}$,
with every tenth such exposure complemented by one in the $V$ band.

The event, which had an
effective duration of only about one day, was essentially over at the
time of the alert.
Moreover, the first pipeline pySIS light curve 
\citep{albrow09} was not posted to the
KMTNet webpage until about five hours later due to a backlog of
%bbb: Hence, no followup observations were possible -> Hence, no followup observation was possible 
newly alerted events. Hence, no followup observations were possible.

Although many dozens of events, including other FSPL events,
 were actively modeled in real time during the 2019 season by up to a 
dozen modelers, this event was not. However,
the online light curve showed clear deviations that are indicative
of finite-source effects in an $A_\max \sim 2$ event on a cataloged source
with dereddened apparent magnitude $I_0 = 14.38$, implying that 
$\rho \sim \sqrt{A_\max^2 -1}/2\sim 1$, and hence
$\theta_\e\sim \theta_* \sim 6\,\muas$ (assuming, as proves to be the
case, that the source color is similar to the clump).
That is, even in the real-time data, this was a very plausible FFP candidate.
While this is only one case, it may indicate that there are other
FFP candidates that are not being recognized from pipeline data
using current human-review and machine-review techniques.

The event was first noticed as a potential FFP candidate in February 2020, 
during
a routine inspection of all events found by the KMTNet EventFinder
\citep{eventfinder}, which included the rediscovery of KMT-2019-BLG-2073.
The data were then rereduced using a tender loving care (TLC) 
implementation of the same pySIS algorithm \citep{albrow09}, which is
a specific variant of difference image analysis 
(DIA, \citealt{tomaney96,alard98}).

\section{{Light Curve Analysis}
\label{sec:anal}}

\begin{figure}
\plotone{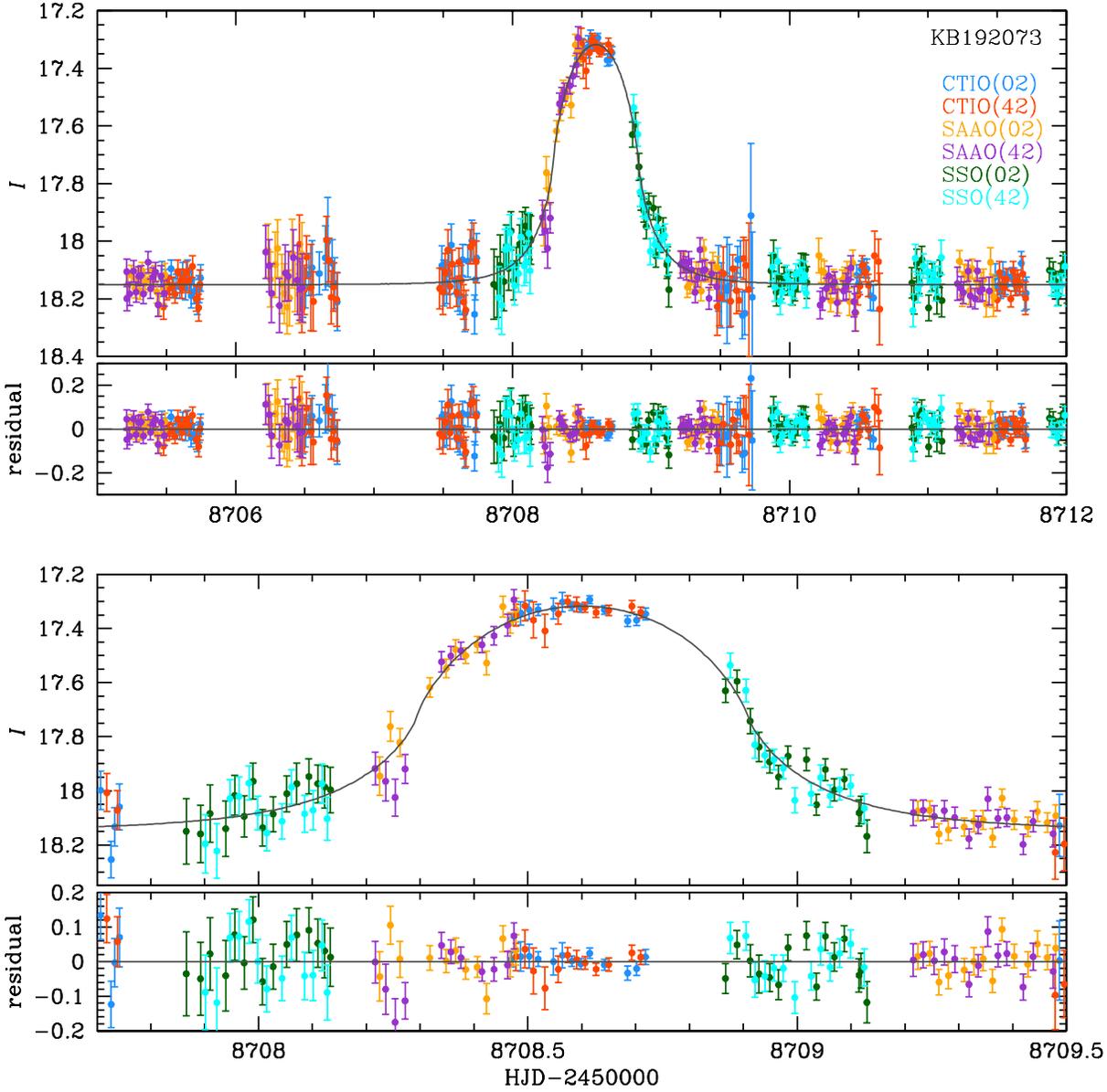}
\caption{Light curve and model for KMT-2019-BLG-2073, which was observed
continuously (weather permitting) at a cadence of $\Gamma=4\,{\rm hr}^{-1}$
from three KMTNet observatories (KMTA, KMTC,KMTS) in two overlapping
fields (BLG02, BLG42).  The upper panel shows a five-day interval containing
the brief event, while the lower panel shows a zoom.  In fact the Einstein
radius crossing time is only $t_\e\simeq 0.31$ days, but the event is stretched
by pronounced finite-source effects, which enable a measurement of the
Einstein radius: $\theta_\e = 4.8\pm 0.2\,\muas$.  This implies a lens
mass of $M=59\,M_\oplus(\pi_\rel/16\,\muas)^{-1}$ where $\pi_\rel$ is the 
lens-source relative parallax.  This planetary lens has no known host.
Each panel also shows residuals.}
\label{fig:lc}
\end{figure}

The light curve appears to be a featureless, symmetric, short-duration
bump that is qualitatively consistent with single-lens/single-source
(1L1S) microlensing.  As mentioned in Section~\ref{sec:obs}, it shows
strong deviations from a simple \citet{pac86} fit, whose
geometry is characterized by three parameters $(t_0,u_0,t_\e)$, i.e.,
the time of peak magnification, the impact parameter (scaled to $\theta_\e$),
and the Einstein timescale.  These deviations are well explained
by adding $\rho=\theta_*/\theta_\e$ as a fourth parameter. In addition,
there are two parameters $(f_s,f_b)_i$ for each observatory $i$ that
represent the source and blended flux, respectively.
See Figure~\ref{fig:lc}.  In deriving this fit, we adopt a linear
limb-darkening parameter $\Gamma=0.53$ based on the source typing
(clump giant) given in Section~\ref{sec:cmd}.
%                     V     I     Gamma
%u 2 3.50 4500 0.0 0.8078 0.6263  0.737   0.528

Table~\ref{tab:single} shows the results of two fits to the data, one with
a free blending parameter $f_b$ and the other with fixed $f_b=0$.
These results show that the free fit is consistent with $f_b=0$
at $1\,\sigma$.  As discussed by \citet{ob190551}, the blending
fractions of $\eta=f_b/f_{\rm base}$ of (apparently) clump-giant sources
are bimodal between values close to 0 and 1.  Because the result
is consistent with $\eta=0$, we will eventually adopt the fixed $f_b=0$
solution.  Nevertheless, in order to better understand the implications
of this choice, we first carefully examine the free-blending solution.

As shown in Table~\ref{tab:single}, for the free-blending fit,
the errors in $t_\e$ and $\rho$ are 10\% and 12\%, 
respectively. However, these are highly anti-correlated, so that
the source crossing time $t_*\equiv \rho t_\e$ is measured
to better than 4\%.  The fractional error in $f_s$ is even larger, 
22\%.  However, we also show in Table~\ref{tab:single},  
the normalized surface
brightness,
\begin{equation}
\hat S \equiv {f_s\over\rho^2} = \pi\theta_\e^2 {f_s\over \pi \theta_*^2},
\label{eqn:Sdef}
\end{equation}
which, like $t_*$, has a much smaller fraction error than either of the factors
that enter it, i.e., $\sigma(\ln\hat S)= 5\%$.  
This parameter combination is motivated 
by the argument of \citet{ob190551}, who showed that in the limit of
$\rho\gg 1$, $\theta_\e=\theta_*/\rho$ is much better determined than
either $\theta_*$ or $\rho$.  The underlying physical reason is that
the excess flux due to a lens acting on a very large ($\rightarrow\infty$)
source of uniform surface brightness $S$, is just 
$\Delta F = 2\pi\theta_\e^2 S$.
Hence, if the source color (and so surface brightness) is considered known,
then the empirically observed excess flux directly gives the Einstein
radius $\theta_\e$, even if $f_s$ (and therefore $\theta_*$) and $\rho$ are
poorly measured.

In the present case, we are not in the limit $\rho\gg 1$, but the same
argument applies reasonably well.  That is, if we assume that the source
color is known, then $\theta_*$ scales with $\sqrt{f_s}$ via,
\begin{equation}
\theta_* = \sqrt{f_s\over f_{s,\rm fid}}\theta_{*,\rm fid},
\label{eqn:thetastarfid}
\end{equation}
where $\theta_{*,\rm fid}$ is the value of $\theta_*$ at some arbitrarily
chosen value of $f_{s,\rm fid}$.  Then, from the measurement of $\hat S$
(Equation~(\ref{eqn:Sdef})),
\begin{equation}
\theta_\e = {\theta_*\over\rho} = 
{\theta_{*,\rm fid}\over \sqrt{f_{s,\rm fid}}}{\sqrt{f_s}\over\rho} = 
{\theta_{*,\rm fid}\over \sqrt{f_{s,\rm fid}}}\hat S .
\label{eqn:useS}
\end{equation}

The source flux $f_s$ is consistent with the baseline flux 
$f_{\rm base} \equiv f_s + f_b$ at $\sim 0.4\,\sigma$ (keeping in mind
that the errors in $f_s$ and $f_b$ are almost perfectly anti-correlated).
Given that the source lies in or near the clump, it is therefore quite
plausible that $f_s$ is essentially equal to $f_{\rm base}$, and that
the apparent difference is due to modest statistical errors.  As we have
just shown, this would make essentially no difference for the $\theta_\e$ 
determination, which is of primary interest.  However, it would affect
the proper-motion estimate, which is also of importance.  That is,
$\mu_\rel = \theta_\e/t_\e = (\theta_\e/t_*)\rho$.  Because $\theta_\e$ and $t_*$
are nearly invariant, $\mu_\rel\propto\rho$.  Thus, the principal difference
between the $f_b=0$ and free $f_b$ solutions is that that latter have
an additional fractional error in $\mu_\rel$
equal to that of $\rho$, i.e., $\sim 12\%$.
As stated above, we adopt the $f_b=0$ solution because of the
low prior probability of intermediate blending parameter $\eta$.  
Nevertheless, for applications that depend sensitively on the proper
motion (e.g., delay time until future high-resolution imaging), this
potential source of proper-motion uncertainty should be given due weight.

\section{{Color Magnitude Diagram}
\label{sec:cmd}}

\subsection{{Overview}
\label{sec:overview}}

For essentially all microlensing events, the main goal of the 
color-magnitude diagram (CMD) analysis is to measure $\theta_*$ and so
$\theta_\e = \theta_*/\rho$ and $\mu_\rel = \theta_\e/t_\e$.  This
is also true in the present case, but with somewhat different emphasis.
First, the $\theta_\e$ measurement is of overarching importance.
Second, in contrast to typical events, this measurement depends almost
entirely on the source color (see Equation~(\ref{eqn:useS}) and 
Table~\ref{tab:single}).  Third, this source color cannot be reliably
measured from the light curve.  Thus, most of this section is focused
on quantifying the uncertainty of the source color and the impact
of this uncertainty on $\theta_\e$ (and also $\mu_\rel$).   However,
in order not to distract the reader with the details of this
investigation, we note at the outset that Equation~(\ref{eqn:evals1}), below,
gives our best estimate of the value and statistical error in $\theta_\e$,
and that the probability of significant systematic deviation from
this result is small.

\subsection{{Source Color Not Measured From Event}
\label{sec:nosourcecolor}}

In order to estimate $\theta_\e$, we find the source's position relative to the
clump on the CMD \citep{ob03262}.  Normally,
the source color would be derived from observations of the event
in two different bands, either by fitting both light curves to the
same model or by regression.  However, this is not possible in the 
present case.  Despite the brevity of the event, there are four
$V$-band measurements over the peak from KMTC.  However, the
extinction (estimated below at $A_I\sim 3.7$) and intrinsically
red source imply that the $V$-band difference fluxes at peak
correspond to a roughly $V\sim 22$ ``difference star''.  This was
too faint for the observations to yield a measurement of useful
precision.

\subsection{{Analysis Assuming That Source = ``Baseline Object''}
\label{sec:source=base}}

\begin{figure}
\plotone{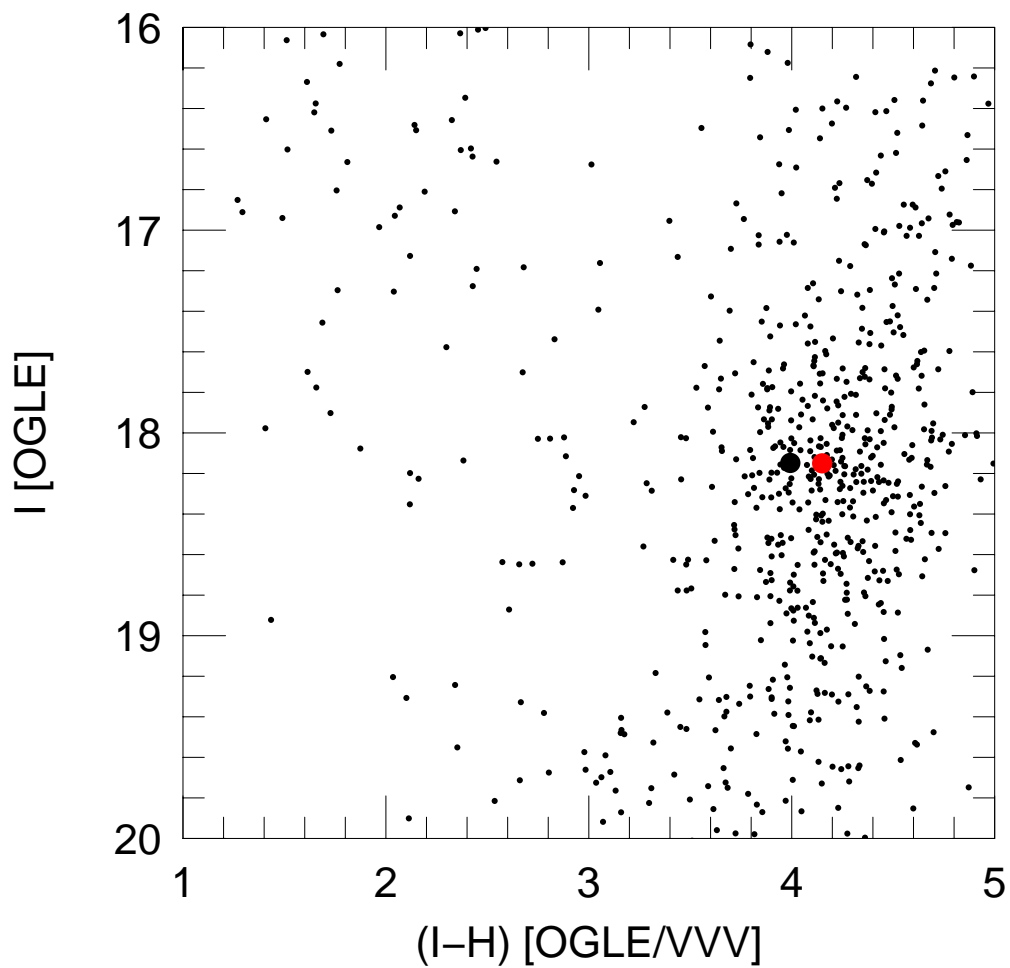}
\caption{Calibrated $I$ vs $(I-H)$ CMD for KMT-2019-BLG-2073 derived
by matching OGLE-III $I$ and VVV $H$ photometry.  The black point shows
the position of the ``baseline object'' derived from these catalogs, while
the red point shows the clump centroid.  We derive the source radius $\theta_*$
under the assumption that it is unblended (source = baseline) and under
the assumption that it is blended.  In either case, $\theta_*\sim 5\,\muas$,
implying that the Einstein radius is $\theta_\e\sim 5\,\muas$ as well.}
\label{fig:cmd}
\end{figure}

Instead, we evaluate the color of the ``baseline object'' using archival
$I$-band photometry from the OGLE-III survey
\citep{oiiicat} matched to archival
$H$-band photometry from the VVV survey \citep{vvvcat} to make a 
first estimate of the source color.  See Figure~\ref{fig:cmd}.

%aaa: why this is approach is appropriate -> why this approach is appropriate
There are two reasons why this approach is appropriate.  First,
we find that the source position, as determined from astrometry on
the difference images from the peak night at KMTC, is aligned within
0.063 pixels (25 mas) of position of the baseline object on the reference image.
Hence, if any star other than the source contributes
significantly to the baseline object, it must be 
very closely aligned with the source.  While this
could certainly be the case for a star associated with the event
(i.e., a companion to the source or a host of the lens), 
it is very improbable for a random interloper.  That is,
%bbb: that are, e.g., 10\% as bright -> that are, e.g., 10\%, as bright
the surface density of stars that are, e.g., 10\% as bright as a clump
giant, is low, so that the probability of one falling within a few tens of
mas of a randomly chosen location is extremely low.

Second, the flux parameters shown in Table~\ref{tab:single} 
for free blending imply that
most of the baseline-object flux is due to the source, and 
the baseline flux is
consistent with all being due to the source.

%0.94   2.16    0.085 1.135
%0.94                 1.134
%1.00   2.31    0.095 1.215
%1.06                 1.294
%1.08   2.50    0.10  1.32

Hence we adopt this $f_b=0$ assumption to make our first estimate for
the source color and magnitude.  According to Figure~\ref{fig:cmd}, the
baseline object (and hence source, under this hypothesis) is separated from
the clump by
$\Delta[(I-H),I] = (-0.16,0.00)\pm (0.02,0.03)$.
Using the dereddened clump position $[(V-I),I]_{\rm clump,0} = (1.06,14.45)$
\citep{bensby13,nataf13} and the color-color relations of \citet{bb88},
this implies
$\Delta[(V-I),I] = (-0.12,0.00)\pm (0.02,0.03)$ and so
$[(V-I),I]_{s,0} = (0.94,14.45)\pm (0.02,0.03)$.  Then, again using the
color-color relations of \citet{bb88} as well as the color/surface-brightness
relations of \citet{kervella04}, we obtain,
\begin{equation}
\theta_* = 5.43 \pm 0.17\,\muas;
\quad
\theta_\e = 4.77 \pm 0.19\,\muas;
\quad
\mu_\rel = 6.41\pm 0.24\,\masyr;
\quad
(f_b=0),
\label{eqn:evals1}
\end{equation}
where we have employed the ``$f_b=0$'' solution from Table~\ref{tab:single}, 
as is appropriate for these assumptions.

\subsection{{Possibility That Source $\not=$ ``Baseline Object''}
\label{sec:notbaseline}}

We now consider the possibility that some of the baseline flux is due
to another star, i.e., other than the microlensed source.  As we noted 
above, the close astrometric alignment of the source with the baseline
object argues against significant light coming from an ambient star.
It is also unlikely that substantial additional light is contributed
by a companion to the source because it would be subject to the
same heavy extinction as the source.  Moreover, the main possibility
would be a subgiant companion because giant companions would be
extraordinarily rare and main-sequence companions would contribute hardly
any light.  In this context, we note that the subgiant would contribute
a modest amount of flux, which would be of similar color to the baseline
object.  Hence, by the argument given in Section~\ref{sec:anal}, the
Einstein-radius estimate would be essentially the same as the
$\theta_\e=4.8\,\muas$ estimate given in Equation~(\ref{eqn:evals1}),
while the proper motion $\mu_\rel\propto \rho$ would be lower by of order 15\%.
Because this solution is qualitatively the same for $\mu_\rel$ and
quantitatively nearly identical for $\theta_\e$, we do not pursue it in detail.

Instead, we examine the possibility that significant blended light comes
from a (putative) host of the lens.  We will derive constraints on this
scenario in Section~\ref{sec:host}, but for present purposes, we simply note
that if a host lay more than a few mas away, it would not have any 
perceptible impact on the light curve, and if it were less than 50 mas,
it would not violate the astrometric constraints.  Thus, this possibility
cannot be ruled out based on existing observations.\footnote{Note that this exercise should {\em not} be mistaken for an estimate of light from a putative host. It simply considers the maximum impact one {\em might} have on the interpretation of the source.}

However, if the putative host lay behind the majority of the
dust, it would face the same arguments as were given above for a
companion to the source.  On the other hand, if the lens were
relatively nearby to the Sun, it could contribute substantial light both
because of its proximity and because it was less extincted than the source.
In this case, it could also contribute significant $V$-band 
light, which would render the source redder than the baseline object.

 From the standpoint of exploring a range of possibilities, this is
the main feature of interest because, according to the argument
of \citet{ob190551}, only a change in the source color can alter
the estimate of $\theta_\e$.  To be concrete, we assume that 
$-8\%\pm 22\%$ of the
$I$-band light is due to the blend (see Table~\ref{tab:single}) 
and that the source color is that of the clump, 
$(V-I)_0 =1.06\pm 0.10$, which is
the most likely color in this region of the CMD.  This could in principle
be achieved if the lens (and so lens companion) lay in theforeground, i.e.,
in front of a substantial fraction of the dust, and so were 
$\Delta(V-I)\sim -1\,$mag bluer than the source.  In any case, it
is highly unlikely that the source can be substantially redder 
than this range because this would put it in a highly underpopulated region
of the CMD.  

 From Equation~(\ref{eqn:useS}), $\theta_\e$ is the product of
two terms, the first depending only on the source color
and second being $\hat S$.  The mean value of the latter does not
change for free blending, although its uncertainty becomes larger, i.e.,
5\%.  When the color is increased from $(V-I)_0 = 0.94$ to $1.06\pm 0.10$,
this increases the color term in Equation~(\ref{eqn:useS}) by
$14\%\pm 11\%$.  Hence, under the assumption that the relatively blue
color (for a clump star) of the baseline object is due to 
``contamination'' by the host of the lens (or a companion of the host),
$\theta_\e$ would be larger by $20\%\pm 12\%$, i.e., $5.72\pm 0.57\,\muas$.
In addition, if, e.g., 10\% of the $I$-band flux were due to the
putative host system, then the proper-motion estimate would be
reduced by 5\% relative to Equation~(\ref{eqn:evals1}).

We conclude that Equation~(\ref{eqn:evals1}) gives the best estimate
of $\theta_\e$ and $\mu_\rel$, but even if the baseline object includes
substantial blue light due to a putative host of the lens that lies in front of
much of the dust, the estimates of $\theta_\e$ and $\mu_\rel$ do not 
qualitatively change.

\subsection{{Comparison of KMT-2019-BLG-2073 and OGLE-2019-BLG-0551}
\label{sec:comp}}

Finally, it is worthwhile to compare how the parameter estimates
of KMT-2019-BLG-2073 and OGLE-2019-BLG-0551 \citep{ob190551} are 
affected by the introduction of free-blending.  For OGLE-2019-BLG-0551,
the color was measured directly from the light curve.  Therefore,
the assumption of fixed color, employed by \citet{ob190551} in their
derivation of the invariance of $\theta_\e$, was fully justified.  
On the other hand, the light curve provided only very weak constraints
on $f_s$ (equivalently, $I_s$).  Therefore, in their Table~1,
$\theta_\e$ is nearly identical between the two models, whereas
$\mu_\rel \propto \sqrt{f_s}$ is much smaller for the free-blend case.
In addition, the fact that the measured color of the event was
the same as the color of the baseline object provided strong
evidence of low blending (although this was not explicitly used
in the mathematical analysis).

By contrast, for KMT-2019-BLG-2073, $f_s$ is well-measured (and
is close to $f_{\rm base}$), while there is no measurement of the source
color per se (only of the baseline object).  Hence, the 
\citet{ob190551} argument for invariance of $\theta_\e$ cannot
be used directly, but must be generalized to Equation~(\ref{eqn:useS}),
which includes a color term.  The color (and hence color term) is 
constrained partly by the morphology of the CMD and partly by the
fact that $f_s$ is measured from the event to be similar to $f_{\rm base}$.
As a result, the difference in the values of $\theta_\e$ are larger
for KMT-2019-BLG-2073 than OGLE-2019-BLG-0551, while those of $\mu_\rel$ are
smaller.

%   rho  te    t*    th*   mu   mutable
%  2.73 0.505  1.38  11.4        3.01

\section{{Search for a Host}
\label{sec:host}}

 Based on its light curve, KMT-2019-BLG-2073 appears to be an
isolated lens of very low mass, i.e., an FFP candidate.  However, if it were
orbiting a star at sufficiently wide separation, then this host
%bbb: added addtional discription, ...a weak trace (because the host may not leave a signal even if it exists)...
would leave only a weak trace of its existence, or perhaps
no trace at all.  In this section, we search for such weak traces
and characterize the separations that we are able to probe.

In our search, we consider models with seven parameters
$(t_0,u_0,t_\e,\rho,s,q,\alpha)$.  The first four are similar
to our 1L1S search, except that $(u_0,t_\e,\rho)$ are all normalized
to the Einstein radius of the combined (binary) lens, which is larger
than the FFP Einstein ring by a factor $\sqrt{q+1}$, where $q>1$
is the host/planet mass ratio.  Then $s$ is the projected host-planet
separation scaled to the host+planet Einstein radius, and $\alpha$
is the angle of the lens-source trajectory relative to the host-planet
axis.  We place the planetary caustic at the center of
our coordinate system, so that $u_0\simeq 0$ and $t_0$ is similar to the value 
listed in Table~\ref{tab:single}.  
We initially conduct a grid search with $(s,q)$ held fixed
and with $\alpha$ seeded at six positions drawn from the unit circle.
We include two years (2018+2019) of data in order to suppress
false signals due to low-level source variability of the giant source.
See \citet{ob190551} for an alternate approach.  

We find no 2L1S models with significant $\chi^2$ improvement.  More 
specifically, after seeding a new fit that is free in all seven parameters
with the best grid point, we find a shallow
minimum at $(s,q,\alpha)=(53,246,12.5^\circ)$, with $\Delta\chi^2=-6.20$ 
relative to the FSPL ($f_b=0$) fit for the same two-year data set and
for four additional degrees of freedom $(s,q,\alpha,f_b)$.  Even if
one were to assume perfect Gaussian statistics (which is certainly not
permissible for microlensing data), this would have a significance
$p = (1+\Delta\chi^2/2)\exp(-\Delta\chi^2/2) = 18\%$, which is not
significant.

Given that there is no significant evidence for the host in the microlensing light curve, we work to place limits on possible hosts by applying the 
Gaudi \& Sackett (2000) method to the full 2018-2019 KMT 
light curves.  As we show below,
this method yields overly strong constraints on hosts because
of low-level variability of the source.  Nevertheless it yields
important insights into the real limits, which then serve to guide
an analytic estimate of the limits.  We implement this method by
conducting
a three-dimensional (3-D) $(s,q,\alpha)$ grid search
in order to put upper limits on the presence of a host. We again
seed the remaining parameters at the FSPL values from Table~\ref{tab:single}, 
The grid is equally spaced in $\log q$ over $1<q<10^6$, equally spaced 
in $\log s$ over $1<s<100$, and uniformly spaced in $\alpha$ in $10^\circ$
steps.  

We find that the constraints derived from this approach are too
strong, due to low-level variability of the source.  For example,
the model $(s,q,\alpha)=(10,100,90^\circ)$ is nominally excluded at
$\Delta\chi^2=14$ with $\Delta\chi^2=10$ coming from the 2018 baseline
year. However, close inspection of the 2L1S and 1L1S models shows
that they only differ by $7\times 10^{-5}$ mag during 
2018\footnote{The baselines systematically differ as 
a natural consequence of the
fact that the 2L1S model has a long $\sim 100\,$day low-amplitude
bump, which is not seen in the data, so the model automatically
depresses the baseline flux to minimize $\chi^2$.}.  
There are about
$N=6000$ points in 2018, with overall median error $\sigma=0.032$ mag.
Hence, the baseline can only be measured to a precision 
$\sigma/\sqrt{N}=4\times 10^{-4}$ mag, so it is clearly impossible
to distinguish these two solutions using these data.  The reason
for the formally high $\Delta\chi^2$ is that the data are about 0.01
mag higher than either model (due to source variability).  This
is a standard systematic effect, which improperly augments the
$\chi^2$ difference by a factor $2\times(0.01)/(7\times 10^{-5}) \sim 300$.
Hence, we cannot blindly apply this approach.

However, studying the models that are generated by this approach, we
find that for each $(s,q)$ in the regime $q>10$, the $\alpha=90^\circ$
model is the most weakly constrained.  For all of these models,
the constraints come from a long low-amplitude ``bump'', which
is not seen in the data.  When $\alpha=90^\circ$, this bump has
equal amplitude on either side of the planetary spike.  At any other
$\alpha$ it is higher on one side or the other, which only increases
its tension with the data.  Therefore, we can restrict attention to
the 2-D parameter space $(s,q,\alpha=90^\circ)$.  We also note
that between the two potential features that are due to a distant host, i.e.,
the long bump and the small caustic at the position of the planet,
only the first induces significant changes in the lightcurve.

In this $\alpha=90^\circ$ geometry and for $s\gg 1$, the separation of
the source and host is given by $u = \sqrt{s^2 + (t-t0)^2/q t_\e^2}$
where $t_\e=0.27\,$days is the Einstein timescale of the FFP solution.
And the excess magnification due to the distant host is well approximated
by $\Delta A = 2/u^4$.  Hence, the FWHM of the bump is given by
$\Delta t_{\rm FWHM} = 2(\sqrt{2^{1/2}-1}s\sqrt{q}t_\e=1.3s\sqrt{q}t_\e$.
That is, 
\begin{equation}
\Delta t_{\rm FWHM} = 78(s/7)\sqrt{q/1000}\,{\rm day}.
\label{eqn:tfwhm}
\end{equation}

Hence, for $q<1000$, both the bump and the adjacent baseline
are contained within the 2019 data, which is relatively unaffected
by the long term variability.  Taking account of the analytic form
of the bump and the photometric errors, we find that for a given $q$
in this regime, $s$ is constrained to be
\begin{equation}
s > 7\biggl({q\over 1000}\biggr)^{1/8};
\qquad 1<\log q < 3.3
\label{eqn:slim_smallq}
\end{equation}
On the other hand for $q>10000$, the
bump extends over most of the 2019 season.  We must therefore
rely on the 2018 baseline measurement, which can only
be constrained to within 0.01 mag due to source variability.
This implies that the limit is $s>3.3$.  That is, for this limiting
value $u=s-1/s \sim 3$ so that $A-1= 0.017$, which can clearly
be distinguished from the baseline, even with the uncertainty
due to variability.  That is,
\begin{equation}
s > 3.3 \qquad \log q > 4.0
\label{eqn:slim_largeq}
\end{equation}
Finally, in the intermediate regime, as $q$ rises, the limiting
value of $s$ must fall $s\propto q^{1/2}$ in order to keep the
bump confined to the 2019 season (according to Equation~(\ref{eqn:tfwhm})),
\begin{equation}
s > 7.6\biggl({q\over 2000}\biggr)^{-1/2};
\qquad 3.3<\log q < 4.0.
\end{equation}
Finally, we note that in the regime $q<10$ (which would correspond
to an exotic FFP planet-moon scenario) the caustic plays
a dominant role and rules out all $s$ in the ranges considered above.

To gain intuition about the physical implications of these constraints, we consider two test cases. Suppose that the lens corresponds to a Jupiter-mass planet. Then, a $1\, M_{\odot}$ host would have $q = 10^{3}$. The physical Einstein radius for such a system is given by
\begin{equation}
r_{\rm E}= \frac{\theta_{\rm E, p} D_{\rm L}}{q}
\end{equation}
where $\theta_{\rm E,p}$ is the angular Einstein radius of the planet, which is $4.8 \mu$as in this case. Equation~(\ref{eqn:meval}) gives $\pi_{\rm rel} = 2.8\ \mu$as and $D_{\rm L} = 7.8\ $ kpc (for $D_{\rm S} = 8$ kpc).
Equation~(\ref{eqn:slim_smallq}) indicates a limit of $s > 7$, which corresponds to a limit of $a_{\perp} = s r_{\rm E} > 8.3$ au. Alternatively, consider the case of a Neptune mass ratio $q = 2\times10^{4}$ and an $0.3\, M_{\odot}$ host. In that case, the limit is given by Equation (\ref{eqn:slim_largeq}), so $s > 3.3$ or $a_{\perp} > 7.1$ au. While these constraints are significant, they would not rule out, e.g., a Neptune analog. We discuss how to place additional constraints on the presence (or absence) of a host star in Section \ref{sec:hostao}.

% april 16 email
%(chi2,t0,u0,te,rho,s,q,alpha)=( 24599.2,8708.6,0.01,1.66,0.19,9.24,45.2,1.396)
% chi2(1L1S)                     24596.8 tab 5
% chi2(2L1S)                     24590.6 tab 6

\section{{Outline of Statistical Approach to Isolated-Lens Finite-Source Events}
\label{sec:stats}}

The detection of two isolated lenses with FFP-class 
($\theta_\e < 10\muas$) Einstein radii in 2019 suggests that
a statistically meaningful number of such lenses could exist
in four years of KMTNet survey data.  This motivates us to develop
a statistical procedure that can guide future searches, including
both archival and prospective data.  Here we present the motivations
for our approach, describe a specific implementation, and then 
apply this implementation to the 2019 KMT database. 

We strongly emphasize that even though this implementation will yield
a statistically well-defined sample, the 2019 sample alone absolutely cannot be used
to derive statistical inferences about FFPs.  This is because, as stated above,
we were motivated to undertake this study due to the detection
of two FFP-class events.  Hence our study is, by definition,
critically impacted by publication bias.  Nevertheless, because
the sample is statistically well-defined, it will allow us to
address possible issues in the construction of a larger, unbiased,
statistical sample. 

Thus, the primary purpose of this work is to establish the methodology for carrying out a statistical study of FFP candidates using the $\theta_\e$ distribution. In  future work \citep[e.g. KB172820 in ][]{kb172820}, we will extend this procedure to the full (unbiased) sample of KMTNet events from which statistical conclusions can be drawn.

\subsection{{Motivation}
\label{sec:motive}}

Our basic goal is to assemble a homogeneous and complete sample of 
1L1S events with secure finite source effects, i.e., FSPL events, in order to characterize the population of FFPs. As mentioned in Section \ref{sec:intro}, the focus on FSPL events has several major advantages over trying to characterize the population based on $t_\e$ alone. 
If, for the moment, we identify such events as those
with $z_0\equiv u_0/\rho \leq 1$, then their underlying cross section
is $2\theta_*$, independent of the mass, distance,
or transverse velocity of the lens.  Hence, (under the assumption
that the distance and proper-motion distributions of the lenses
is independent of the lens mass $M$),  each subclass of lenses
(classified by, e.g., their mass) will contribute to this
homogeneous sample in direct proportion to their number density. This is an advantage over the $t_\e$ distribution, to which lenses contribute in proportion both to their number density and to (the square-root of) their mass.

The second advantage is that, for each member of the sample, the measurement of $\theta_\e$
will imply a probability distribution for the lens mass that is
directly related to the Galactic-model distribution of lens-source
relative parallax, $\pi_\rel$,
\begin{equation}
M = {\theta_\e^2\over\kappa\pi_\rel}.
\label{eqn:mthetae}
\end{equation}
Thus a $\theta_\e$-based sample is subject to much less uncertainty
than a $t_\e$-based sample, for which
$M\propto \mu_\rel^2/\pi_\rel$. Furthermore, there does not appear to be a large population of free-floating Jupiters, but rather the FFP population, should it exist, must peak at lower masses. Hence, 
because the distribution 
of $\pi_\rel^{-1}$ is relatively compact, while the peaks of the planetary
and stellar mass functions are separated by several orders of magnitude,
%aaa: distribution of $\theta_\e^2$, can directly -> distribution of $\theta_\e^2$ can directly 
the observed distribution of $\theta_\e^2$ should also be bimodal and thus,
can directly constrain the relative frequency of stars and FFPs.

However, the key to being able to make a statistical statement lies in having a well-defined and statistically robust sample of events with $\theta_\e$ measurements.
As usual, the devil is in the details of the selection function.
For ``complete selection'', all of the specified class of FSPL
events must be identified as microlensing candidates (not necessarily
immediately identified as FSPL events), and must be closely examined
%aaa: with high-quality reductions for finite-source effects. -> with high-quality reductions for the solid detection of finite-source effects.
with high-quality reductions for the solid detection of
finite-source effects.  That is, they must
be selected independent of $\theta_\e$, or at least with smoothly
varying and reasonably well known selection as a function of $\theta_\e$.

\subsubsection{Advantages of Giant Sources}

These goals are strongly aided by restricting the investigation
to giant-star sources.  There are a number of reasons for this.

First, the fraction of events with
underlying (not necessarily detected) finite-source effects is
%aaa: ..., which is or... -> ..., which is of...
directly proportional to $\theta_*$, which is of order 10 times
larger for giants than dwarfs. Thus, it is more efficient to search for FSPL events with giant sources.

Second, the
machine classification of source-star characteristics is much
more accurate for giant sources because they are typically
much less blended. The first implication of this is that it is easier to create a well-defined sample of giant sources. In addition, for lower-mass lenses,
for which FSPL events are typically fainter at peak, 
so that the true position centroid
of dwarf sources is much less likely to be recovered by automated means. Because the accuracy of the centroid affects the quality of the photometry, 
finite-source event candidates can be more reliably
identified for giant sources.  

Third, because giant sources are brighter, they tend to have higher-precision photometry simply due to improved photon noise. As just mentioned, better photometry improves the detectability (and characterization) of finite-source effects.

Fourth, the number of giant-source events
is an order of magnitude smaller, which makes for a more tractable sample. In addition, it is more likely to be complete than a larger sample. For example, considering dwarf-source events
creates an order of magnitude \HL{more}
work in careful vetting of candidates, while we expect a smaller fraction of them to exhibit finite source effects, and thus increases the
chance that a significant fraction of the finite-source events will
be missed.

Fifth, because giant sources have a larger $\theta_*$, they also have a longer $t_*$ at fixed $\mu_{\rm rel}$, which has a number of consequences. First, the longer duration means that more observations are \HL{taken} that characterize the event at fixed cadence. This makes the giant source, FSPL events both better characterized and easier to detect than either dwarf source FSPL or FFP-candidate PSPL events. By contrast,
$t_*\sim 1\,$hr for dwarf stars, 
whereas the cadence of the survey fields ranges from $4\, {\rm hr}^{-1}$ to $1\,  {\rm day}^{-1}$ or less. As a consequence,
%aaa: sentence changed, ...effects are much sparser and much more frequently non-existent for dwarf sources. -> ...effects are many times fewer (and often completely absent) for dwarf sources.
the data probing the finite-source effects are many times
fewer (and often completely absent) for dwarf sources.

The second, third, and fourth points all make the selection
function simpler and also easier to accurately model
for giant sources compared to dwarfs.

\subsubsection{Limits of Giant Sources}

The fundamental limit of a giant-source FSPL-based search for FFPs is that
at sufficiently small masses, the ratio $\rho\equiv \theta_*/\theta_\e$
rises well above unity, which has two major consequences.

First, when $\rho > 1$,
 the excess magnification
%aaa: added Reference eq(3) and (4), maeder73,riffeser06,agol03
\citep{maeder73,riffeser06,agol03},
\begin{equation}
A-1 = \sqrt{1 + {4\over\rho^2}}-1,
\label{eqn:excessmag}
\end{equation}
scales as
\begin{equation}
A-1 = \rightarrow {2\over\rho^2}
= {2\kappa\pi_\rel\over\theta_*^2}M .
\label{eqn:ascale}
\end{equation}
Thus, the photometric and systematic noise and/or the level of 
source variability (as
well as the larger $\theta_*$) set the fundamental limit on the mass
that can be detected.

Second, the events may no longer look like standard \citet{pac86} light curves. Hence, they may be missed by an event-detection algorithm designed to detect \citet{pac86} curves, such as the one used in the standard KMTNet pipeline. \HL{While} this
practical problem of recognition can be ameliorated because it is known\HL{, it} will need to be studied in detail (see also Section \ref{sec:supplement}).

Finally, we note that because giant sources are bright, it will be difficult to impossible to detect excess light due to a dwarf host star (should one exist) while the source and lens are still super-posed. \HL{While} this difficulty can be overcome simply by waiting for the lens and source to separate sufficiently far that they can be resolved\HL{, the wait time can be quite long.} See Section \ref{sec:discuss}.

Hence, while dwarf sources do have some advantages, including sensitivity
to FFPs of substantially smaller $\theta_\e$ (and so mass $M$),
it is premature to include them in an initial investigation, i.e., before
giant-source events have been thoroughly investigated. Thus, our overall approach is to identify giant-source-star candidates
automatically, and then to vet these by detailed individual investigation.

\subsection{{Specific Implementation}
\label{sec:implement}}

We present a specific implementation of a FSPL search of giant
sources based on KMTNet data.  In principle,
the same general principles outlined in Section~\ref{sec:motive} could
be applied to more complex data sets, involving, for example, several
microlensing surveys.  However, this would significantly increase the
complexity of the search process, as well as modeling the selection
function.  Nevertheless, such an approach could be adopted at a later time.

Each event is selected for further investigation based
on the four catalog parameters: $I_s$, $u_0$, $t_\e$ and $A_I$.
The first three are the source magnitude, the impact parameter and the
Einstein timescale, all derived from the pipeline PSPL fit.  The
last is the $I$-band extinction, which is estimated as $A_I=7 A_K$
where $A_K$ comes from \citet{gonzalez12}.  We determine 
$I_s = 28 - 2.5\log(f_s)$, where $f_s$ is the source flux in KMTC
instrumental (ADU) units.  We find from extended practice that
this is roughly calibrated, within about 0.1 mag.  For the case of
catalog values $u_0<0.001$, we adopt $u_0=0.001$.
Our selection is
based on two criteria, which we first list and then motivate:
\begin{enumerate}
\item[]{(1)\ $I_{s,0}< 16$;\quad $I_{s,0}\equiv I_s - A_I$ }
\item[]{(2)\ $\mu_{\rm thresh} > 1\,\masyr$; \quad $\mu_{\rm thresh} \equiv 
\theta_{*,\rm est}/t_\eff$; \quad $\theta_{*,\rm est}\equiv  
3\times 10^{(16- I_{s,0})/5}\,\muas$}, 
\end{enumerate}
where $t_\eff\equiv u_0 t_\e$.

Criterion (1) is simply that the source is a giant
(i.e., it has reached the base of the giant
branch in its evolution, which is roughly 1.5 mag below the clump).  In 
principle, stars satisfying this criterion could be foreground main-sequence
stars.  However, if so, these would be eliminated at a later stage.

Criterion (2) is more complex.  The overall goal is to eliminate the great
majority of giant-source events from consideration, while still 
preserving essentially all those with FSPL effects and
proper motions $\mu_\rel\ga 1\,\masyr$.  Because a very small fraction of 
microlensing events have $\mu_\rel< 1\,\masyr$, the overwhelming majority of FSPL
events will survive Criterion (2).  Specifically,
the underlying idea is that 
$\mu_{\rm thresh}$ would be the estimated lens-source relative 
proper motion for the case $z_0\equiv u_0/\rho=1$.  We note
first that 
$\mu=\theta_*/t_* = \theta_*(u_0/\rho)/(u_0 t_\e) = z_0\theta_*/t_\eff$.
Therefore,
$\mu=z_0\mu_{\rm thresh}(\theta_*/\theta_{*,\rm est})$.  Hence, if finite source
effects are detectable, then $z_0\la 1$, and thus events that fail Criterion (2)
would have $\mu \la (\theta_*/\theta_{*,\rm est})\,\masyr$.
The estimate of $\theta_*$ is based on the assumption that the
source has approximately the color of the clump.  If this is approximately
correct, then any event with detectable finite source effects 
that failed criterion (2) would have proper motion
$\mu\la 1\,\masyr$.  Such events are 
very rare.  For a very small fraction of sources, 
the source may prove to be substantially redder than the clump, in 
which case, it could be, e.g.,  that $\theta_*\sim 2\theta_{*,\rm est}$.
Then some events with proper motions as high as $\mu\sim 2\,\masyr$
could be eliminated automatically, i.e., prior to human review.  
These are also relatively rare.  In any event,
this is a well defined, objective criterion.  Hence it eliminates
events in a deterministic way that can be modeled, if necessary.
The purpose of this criterion is to avoid detailed investigations
of events with very low probability of having detectable finite-source
effects.  We will examine the efficacy of the ``$1\,\masyr$'' boundary
on selection in Section~\ref{sec:apply}.  Finally, as noted above,
the fact that events with $z_0\la 1$ will generally 
have detectable finite-source effects implies that events with 
%ccc: mu_{\rm tresh} -> mu_{\rm thresh}
$\mu_\rel\la\mu_{\rm thresh}$ will have such effects.  This means that events
with $\mu_{\rm thresh}\ga 7\,\masyr$ are excellent FSPL candidates because
most microlensing events have proper motions below this threshold.
We will specifically test this idea in Section~\ref{sec:apply}.

After an event is selected, it is fit both with (FSPL) and without
(PSPL) finite-source effects using the final pipeline pySIS reductions,
and the $\Delta\chi^2 = \chi^2({\rm PSPL}) - \chi^2({\rm FSPL})$ is noted.
If $\Delta\chi^2>15$, then it is accepted as a FSPL event, and if
$\Delta\chi^2<3$, then it is rejected.  For the remaining few cases, we
made TLC re-reductions.  Reanalysis then decisively resolved into
FSPL ($\Delta\chi^2>20$) or PSPL ($\Delta\chi^2<3$) for all of these events.

\subsubsection{{AlertFinder and EventFinder Events}
\label{sec:AFEF}}

We begin by considering all KMT events discovered during 
the 2019 season, either by the KMT AlertFinder \citep{alertfinder}
in real time or the KMT EventFinder \citep{eventfinder} in post-season
analysis.  Comparison of these two samples shows that 581 AlertFinder
events were not recovered by the EventFinder.  While many of these
were spurious or very low-quality events, many others are clearly real,
and therefore were missed either because they were excluded by the
automatic selection of candidates or were misclassified as ``not microlensing''
by the operator\footnote{There were also 969 EventFinder events that
were missed by AlertFinder.  However, this shortfall is to be expected because
AlertFinder did not operate in the wings of the 2019 season, does not
fit data from the falling part of the light curve, and does not
simultaneously fit data from overlapping fields.}.  

\subsubsection{{Supplemental Search}
\label{sec:supplement}}

To test for (and possibly find) additional FSPL giant-star events
that were missed by both AlertFinder and EventFinder, we conduct an
additional search based on a modified version of EventFinder.  
Although, we will apply this search to the 2019 sample, we did not
expect (and do not find, see below) many new FSPL events in the 2019
data.  Rather, this supplemental search was created in the context
of the long term goal of creating a homogeneous sample from four
years of survey data, for which the original search algorithms evolved
over time. 

Hence, in order to both motivate and explain this search, we very 
briefly review the key features of EventFinder and its evolution.  
All light curves are modeled by a grid
of several thousand 2-parameter ($t_0,t_\eff$) \citet{gould96} 
models\footnote{In fact, \citet{eventfinder} fit to two variants
of 2-parameter models, the original \citet{gould96} ``high magnification''
$(u_0 = 0)$ model and a second ``low-magnification''  $(u_0 = 1)$ model.
\citet{eventfinder} showed that for perfect data, these provide remarkably
similar fits over the $\pm 5\,t_\eff$ baselines that they modeled.
However, these two variants can differ in their response to imperfect data, 
leading to different $\Delta\chi^2$ and, more importantly, different automated
displays.  This will be important further below.}
that are restricted to a $|t-t_0|<5\,t_\eff$ baseline.  The 
$\Delta\chi^2$ difference between this fit and a constant model
is noted, and the model with the highest such $\Delta\chi^2$ is
selected for this event.  If this model survives the elimination
of the highest $\Delta\chi^2$ point from each observatory, 
%aaa: ...>\Delta\chi^2_{\rm gould,min}$ it is...  -> ...>\Delta\chi^2_{\rm gould,min}$, it is...
and if it exceeds a certain threshold $\Delta\chi^2_{\rm gould}
>\Delta\chi^2_{\rm gould,min}$, it is written to a file.  

Before discussing how this file is further processed, it is important to 
note that beginning ``halfway'' through\footnote{As discussed below,
about 60\% of the catalog stars come from the OGLE-III catalog \citep{oiiicat}
and the great majority of the remainder come from the DECam catalog
\citep{decam}.  The changeover occurred in 2017, after processing the
OGLE-III stars and before processing the DECam stars.}
the 2017 EventFinder analysis, 
this procedure was modified to add a second
step.  Events that pass the $\Delta\chi^2_{\rm gould}$threshold are
fitted to a 3-parameter \citet{pac86} model, and must similarly
exceed a $\Delta\chi^2_{\rm paczynski}>\Delta\chi^2_{\rm paczynski,min}$ 
threshold.  At first
sight this seems more restrictive, but for 2015-2017a, 
$\Delta\chi^2_{\rm gould,min} = 1000$, whereas for 2017b-2019,
$\Delta\chi^2_{\rm gould,min} = 400$ and $\Delta\chi^2_{\rm paczynski,min} = 500$. The
additional Paczy\'nski test had the effect of finding lower-signal
%aaa: events while at the same eliminating -> events while at the same time eliminating
events while at the same time eliminating a much larger fraction of
low-signal spurious candidates that would have required manual
rejection by the operator.  This also enabled the search to include
events with effective timescales $t_\eff \geq (3/4)^4 (=0.3164)\,$days, whereas
the previous approach was limited to $t_\eff \geq 1.0\,$day.

The next step is to group similar candidates (as determined by
their values of $t_0$, $t_\eff$ and angular position) into groups
by a friends-of-friends algorithm.  Only the ``group leader''
(as determined by $\Delta\chi^2$) is further considered.
This ``group leader'' is first
vetted against a list of known variables and artifacts, and, if it passes,
it is shown to the operator.

For the modified EventFinder, we first restrict attention to potential
giant sources, defined by $I_{\rm cat}-A_I<16.2$, where $I_{\rm cat}$ is the
input-catalog ``$I$-band'' magnitude.  Wherever possible, the input catalog
is derived from the OGLE-III star catalog \citep{oiiicat}, which
is on the standard Cousins system.  Nearly all the remaining
catalog entries (about 40\%) are derived from the \cite{decam} catalog
%aaa: For these, case... -> For these case, ...
based on DECam data.  For these cases, $I_{\rm cat}$ is the catalog value
of the SDSS $i$ magnitude\footnote{When there are no $i$-band source fluxes
tabulated by \citet{decam}, a more complex procedure is applied to
estimate $I_{\rm cat}$.  However, a giant would have to suffer
extreme extinction, and hence have nearly unusable photometry,
to be lacking an $i$ measurement.}.  This value is similar to Cousins $I$
for low- or moderately-extincted red giants, but is a few tenths
higher (``fainter'') for heavily extincted giants because the SDSS
$i$ bandpass is bluer than Cousins $I$.  Hence, this could in 
principle reject some sources that have $I<16$.  However, we will
show in Section~\ref{sec:apply} that this is a small effect.

Finally, a small fraction of catalog entries that lie in regions
not covered by either the OGLE-III or DECam catalogs derive
from DoPhot \citep{dophot} photometry of KMT images.  This
photometry is aligned to the OGLE-III catalog to within about 0.1 mag
and so is on the Cousins scale.

In addition to restricting the catalog stars to giants, we also
restrict the models to those with effective timescales of $t_\eff < 5\,$days.
This is a conservative cut because most longer $t_\eff$ events would
be rejected by Criterion (2) above.  Moreover, those giant-source
events that fail this criterion would be very obvious candidates in
the regular EventFinder search and so would be unlikely to be missed.

The giant catalog stars selected in this way are then
subjected to a \citet{gould96} search with $\Delta\chi^2_{\rm gould,min}=1000$
and are not subjected to further \citet{pac86} vetting.  In this
sense, the search resembles those from 2015-2017a.  This feature
will eventually enable reasonably homogeneous integration of 
2016-2017a EventFinder searches into a full statistical search
at a later time.  However, in contrast to these early searches,
it is carried out for effective timescales $t_\eff \geq 0.3164\,$days
rather than 1.0 day.

Finally, to enhance the operator's ability to spot non-standard events,
in particular those with very large finite-source effects, each event
is displayed with three fitting panels (for ``$u_0=0$'', ``$u_0=1$'', and 
``Pacy\'nski'' fits) rather than one panel 
(for best of ``$u_0=0$'' and ``$u_0=1$'' fits).  Such multiple displays
would be of little benefit for relatively long events that roughly 
approximate a \citet{pac86} model.  However, for short events that are
dominated by non-standard features, and possibly short-lived deviations
due to systematics, the three displays can be quite 
different\footnote{In fact, while FFP candidate 
OGLE-2019-BLG-0551 \citep{ob190551} was recovered by AlertFinder 
as KMT-2019-BLG-0519, it was not recovered by EventFinder for
two distinct reasons.  First, it failed the $\Delta\chi^2_{\rm paczynski}>500$
test (despite having $\Delta\chi^2_{\rm gould}>2000$).  This was the
motivation for eliminating the $\Delta\chi^2_{\rm paczynski}$ cut from the
giant-star special search.  But, in addition, the display of the best
(i.e., ``$u_0=0$'') \citet{gould96} fit really does not look like
microlensing, and was rejected by the operator despite relaxed standards
for the initial trial of the
giant-source special search.  However, the displays derived from
both the ``$u_0=1$'' \citet{gould96} fit and, especially, the \citet{pac86}
fit both look like ``obvious microlensing''.  This motivated the expanded
display.}.

For 2019 data, one does not expect to find many new FSPL events from
this additional search.  In particular, most of the 2019 season data have
already been searched twice, with AlertFinder and EventFinder, providing
some protection against problems and operator error in either search.
However, when the FFP
study is extended to earlier years, we expect that it may find
some very short events, including perhaps FFP candidates that
were missed previously due to the higher $t_\eff$ threshold, as well
as events like OGLE-2019-BLG-0551 that was missed by the normal
EventFinder search in 2019, even though it was found in a separate
(AlertFinder) search.

In fact, this supplemental 2019 giant-source EventFinder search finds
a total of 254 candidates of which 18 are ``new'' (i.e., not
found in the regular EventFinder or AlertFinder searches).  Applying
Criteria (1) and (2) to these 18 ``new'' events yields
one candidate for further consideration, which proves not to 
exhibit finite source effects. This candidate is designated KBS0111, where ``S" stands for ``supplemental".
%EFB190111 %ob190268
This confirms our general expectation,
above, that the supplemental search would not yield many additional
finite-source events for 2019.

On the other hand, when we apply Criteria (1) and (2) to the remaining
sample of $(254-18=236)$ events, we find that we recover 27 of the 
40 candidates found by EventFinder + AlertFinder, including 11 of the 13
with detectable finite-source effects (see Section~\ref{sec:apply}).  These 
11 include both FFP candidates (OGLE-2019-BLG-0551 and KMT-2019-BLG-2073).
The two finite-source events that were not found in the supplemental
search both failed the $t_\eff < 5\,$day criterion, which was included
because these are expected to easily be found by EventFinder.  
In addition, it recovered two events that were not selected based
the EventFinder and/or AlertFinder detections, due to incorrect
pipeline PSPL fits.  
%aaa: changed by professor, One of these... -> Neither of these...
%One of these is likely to be a 2L1S event and the other does not 
Neither of these have detectable finite-source effects.
%OB190382 and OB190268
This
shows that the supplemental search is a powerful check, which will
be important for the analysis of previous years when AlertFinder
was either not operating, or operating in very restricted mode.

It is also true that for 2019, there could be EventFinder events
with $400 < \Delta\chi^2_{\rm gould} < 1000$.  In Section~\ref{sec:apply},
we will show that this is a minor effect.  The decision on how to
handle such events must be made when a rigorous multi-season analysis
is carried out.  For the present, we include events that are identified
in any of the three searches.

\subsection{{Application to KMTNet 2019 Season}
\label{sec:apply}}

\subsubsection{{Final Sample}
\label{sec:finalsample}}

For its 2019 season, KMTNet found more than 3300 candidate microlensing events,
from which we eventually identified 13 giant-source FSPL events,
i.e., smaller by a factor 250.  Here,
we describe this down-selection in detail.  We emphasize that machine
selection (using Criteria (1) and (2)) reduced the original sample by
a factor 60, which is what made detailed human review of the remaining
sample feasible.

Machine application of Criteria (1) and (2) to the three searches
(EventFinder, AlertFinder, and supplemental) yielded a list of
%a total of 53 potential candidates.  Of these, 10 were eliminated
% + EFB0111 nofs + EFB0021/ob190382 2L1S? + EFB0044/ob190268 nofs
a total of 56 potential candidates.  Of these, 10 were eliminated
by visual inspection.  Two (KB192781, KB192863)
had saturated photometry and so could not
be analyzed, while the reductions were poor and unrecoverable for
one other (KB190578).  Rereduced data showed that two 
(KB192222, KB193100) of the 10 are not
microlensing.  Three (KB191841, KB192322, KB191420)
have insufficient data near peak to be analyzed
for finite-source effects (the last because it occurred at
the end of the season).  One (KB192530)
appears to be a highly-extincted giant in 
the automated search but is actually a foreground dwarf.
And one (KB193100) is a potentially interesting microlensed variable,
but cannot be analyzed in the present context.

%aaa: changed by professor, fore -> three (due to remove ob190382)
We further eliminate three events that are double-lens/single-source (2L1S),
%We further eliminate four events that are double-lens/single-source (2L1S),
or possibly 1L2S,
rather than 1L1S.  KMT-2019-BLG-2084 may actually be a ``buried planet''
event (like MOA-2007-BLG-400, \citealt{mb07400}), although there is another
$q\sim 1$ solution.  It is currently
under investigation (Zang et al., in prep). 
OGLE-2019-BLG-0304 (KMT-2019-BLG-2583) has
a low-amplitude second ``bump'' about 70 days after the main peak,
which is almost certainly due to a second lens or possibly second source.
MOA-2019-BLG-256 (KMT-2019-BLG-1241) is a binary lens, very likely
composed of two brown dwarfs \citep{mb19256}. 
%aaa: removing OGLE-2019-BLG-0382 (EFB0021) is plausibly a 2L1S event, although this will require further investigation.
%aaa: Accoding to Han's comment, (1) the brief signal at HJD'~8569.5 turned out to be fake, and (2) there is no FSFP signal
%OGLE-2019-BLG-0382 (EFB0021) is plausibly a 2L1S event, although this will require further investigation.

%This leaves 40 events, which we show in Table~\ref{tab:selected} 
%aaa: changed by professor, 42 events in ms2.tex -> 43 events in ms3.tex
This leaves 43 events, which we show in Table~\ref{tab:selected} 
ranked inversely
by the selection parameter, $\mu_{\rm thresh}$.  The table contains
the four input parameters $(u_0,t_\e,I_s,A_I)$ from the KMT web-based
catalog, the derived parameters $(I_{s,0},\mu_{\rm thresh})$, the value
of $\chi^2_{\rm gould}$ from the EventFinder program, and a field
indicating whether or not the normalized source size, $\rho$, could be measured
(i.e., measurable finite-source effects).  It also contains the
discovery name of the event as well as the KMT name, which are the
%aaa: changed by professor, 42 events in ms2.tex -> 43 events in ms3.tex
same for 25 out of the 43 events.

The ordering of the table shows that $\mu_{\rm thresh}$ is a powerful method
of identifying FSPL candidates.  Detailed analysis of individual events
shows that all of the first nine ($\mu_{\rm thresh}> 6.8\,\masyr$) have $\rho$
measurements, none of the final 15 ($\mu_{\rm thresh}< 1.64\,\masyr$)
%aaa: changed by professor, four of the 18 -> four of the 19
have them, and four of the 19 in between these limits have $\rho$
measurements.  The fact that there is only one $\rho$ measurement
for $\mu_{\rm thresh}<2.3\,\masyr$ and none below $1.64\,\masyr$ strongly 
suggests that very few FSPL events are lost by Criterion (2).

Another notable feature of Table~\ref{tab:selected} is that there are only three
events with $\Delta\chi^2_{\rm gould} < 1000$, namely
KMT-2019-BLG-2528, KMT-2019-BLG-1477, and KMT-2019-BLG-2220.
The first of these is a special case.  The event is almost completely
confined to the month of ``pre-season data'', which were taken for a subset
of western fields, only by KMTC, and only in $I$-band.  This means, first,
that the same event occurring slightly later would have had 
$\Delta\chi^2_{\rm gould}> 1000$ simply because there would have been
data from all observatories.  And second,  the color estimate
(hence the estimate of $\theta_*$) is more uncertain than most other 
events because the color is not measured from magnified 
data\footnote{However, the situation is qualitatively similar for
KMT-BLG-2019-2073.  See Section~\ref{sec:cmd}.}.  
Neither of the other two events have
measurable $\rho$, nor are they expected to given that
$\mu_{\rm thresh} \leq 1.12\,\masyr$ in both cases.  Thus, the threshold
of $\Delta\chi^2_{\rm gould,min}= 1000$ seems generally sensible, although
it would be valuable to test its role in a larger data set.

Table~\ref{tab:parms} gives the FSPL fit parameters for 11 of 
these 13 events.  For the two FFP candidates, 
OGLE-2019-BLG-0551 (KMT-2019-BLG-0519) and  KMT-2019-BLG-2073, these parameters
are given in \citet{ob190551} and Table~\ref{tab:single} of the current
paper, respectively.

Table~\ref{tab:fspl} gives the values 
of $(\theta_*,\theta_\e,\mu_\rel,\mu_{\rm thresh},z_0)$, 
which can all be inferred from 
Tables~\ref{tab:selected} and \ref{tab:parms}, together with the 
dereddened CMD values $[(V-I),I]_{s,0}$ that are given in 
Table~\ref{tab:fspl}.  These CMD values are mostly derived
by the standard approach that is described in Section~\ref{sec:cmd},
but using KMT $V$ and $I$ magnified data.
The exceptions are described in the comments on individual events in
Section~\ref{sec:notes}.

By construction,
we expect $\mu_\rel \la \mu_{\rm thresh}$,  and this relation holds generally,
except for OGLE-2019-BLG-0551 and KMT-2019-BLG-1143.
These outliers are explained by being exceptionally red sources,
which causes the magnitude-only machine determination of $\theta_{*,\rm est}$
to be underestimated.  However, in these two cases, the excess is modest:
$\mu_\rel/\mu_{\rm thresh}= 1.21$ and 1.23, respectively.  Recall from
Table~\ref{tab:selected} that there were no events with detectable
finite-source effects with $\mu_{\rm thresh}/\mu_{\rm limit}<1.64$, where
$\mu_{\rm limit}=1\,\masyr$ was the selection limit.
This comparison again emphasizes the generally
conservative character of Criterion (2).

\subsubsection{{Sanity Checks}
\label{sec:sanity}}

Understanding the selection function (or detection efficiency) to events is critical for statistical characterization of the FFP mass function based on any sample. As argued in Sections \ref{sec:intro} and \ref{sec:motive}, one of the major advantages of a $\theta_\e$-based characterization is that the selection function is largely independent of lens mass. By contrast, the selection function for a $t_\e$-based characterization is strongly dependent on mass, and so requires detailed injection-recovery tests in order to reconstruct the underlying distribution. Of course, the selection function of any search may be affected by unanticipated effects, although those will only affect our characterization of the FFP population if those effects depend on the lens mass.

Thus, we now examine several statistical characterizations of this sample,
with the aim of probing for ``irregularities'' in the selection function, whether anticipated
or unanticipated.  For example, the first investigation examines
the cumulative distribution of the impact parameter $u_0$ relative to the
normalized source size $\rho$, which one expects to be uniform.  But other
investigations are more open ended.

\begin{figure}
%ccc: path change
%\plotone{../eventfinder/z.eps}
\plotone{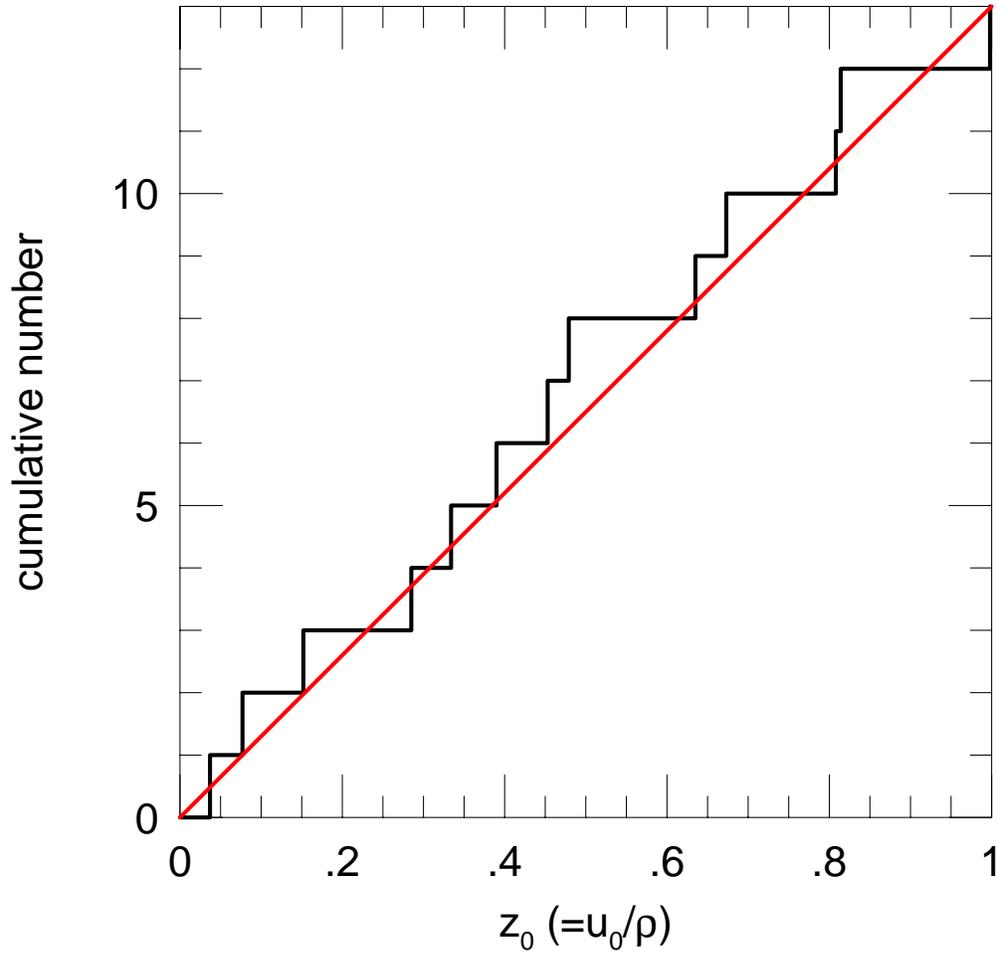}
\caption{The cumulative distribution of $z_0\equiv u_0/\rho$ is quite
consistent with a straight line, which would be the expected behavior
for a sample without selection biases,  In principle, it could be
more difficult to detect finite-source effects for $z_0\simeq 1$
than lower values because the duration of these effects is shorter.
However, this effect, if present, does not have a noticeable impact.}
\label{fig:z}
\end{figure}

Figure~\ref{fig:z} shows the cumulative distribution of $z_0\equiv u_0/\rho$.
Under the assumption that there is no selection bias over the range
$0\leq z_0 \leq 1$, this should be a straight line.  One may expect that
it is more difficult to detect finite-source effects for $z_0\sim 1$,
which would result in a deficit at these values, causing the
%bbb: This is because $z_0\sim 1$ events would seem to suffer deviations from a standard \citet{pac86} curve for a shorter time, which would both increase the chance -> This is because the duation of deviations from a standard \citet{pac86} curve would be shorter for $z_0\sim 1$ events, causing the increase of the chance
cumulative distribution to flatten.  This is because $z_0\sim 1$
events would seem to suffer deviations from a standard \citet{pac86} curve for
a shorter time, which would both increase the chance that these effects
will be entirely missed due to gaps in the data, 
and reduce the statistical significance
of detections to the extent that the data do capture this region.
Contrary to this naive expectation, 
the cumulative distribution seen in Figure~\ref{fig:z} is consistent
with being uniform over $0<z<1$.

The cause of this robustness may be that finite-source deviations are actually
relatively pronounced for $z\equiv u/\rho\ga 1$.  In the high
magnification limit (which generally applies to most of the events
in this sample), one finds using the hexadecapole approximation
\citep{pejcha09,gould08} that the fractional change in magnification
tends toward
\begin{equation}
{\delta A\over A}\rightarrow {1 - (1/5)\Gamma\over 8 z^2}+
{1 - (11/35)\Gamma\over (64/3) z^4};
\qquad (z>1),
\label{eqn:ztend}
\end{equation}
where $\Gamma$ is the limb-darkening coefficient (see Section~\ref{sec:cmd}).
Equation~(\ref{eqn:ztend}) actually works quite well 
almost to $z\rightarrow 1$ (Figure 3 from \citealt{ob151482}).
Hence, it 
shows that, e.g., at $z=\sqrt{2}$, the deviation is about
$\delta A/A \sim 6.6\%$.  Thus, for $z_0=1$, the deviations induced
by finite-source effects remain at of order this level for $\sim 2 t_*$.
Giant-star events have two advantages in this regard.  First, of course,
$2t_*$ is longer for these events, typically 5 hours or more.  However,
there is also a second, more subtle effect.  In events
for which the lens does not actually transit the source,
$\rho$ is highly degenerate with $u_0$, and for most high-magnification
events, $u_0$ is degenerate with $t_\e$ and other parameters.  Disentangling
these degeneracies requires high quality data near baseline, i.e., 
$A\sim$ few.  However, if the source is very faint, then the
photometric errors near baseline are large compared to $(A-1)f_s$.
But for giant sources, these fractional errors are much
smaller unless the giant happens to be highly extincted.
In brief, Figure~\ref{fig:z} suggests that there is no
strong bias against detection of finite-source effects at $z_0\sim 1$,
and theoretical considerations tend to support this suggestion.

\begin{figure}
\plotone{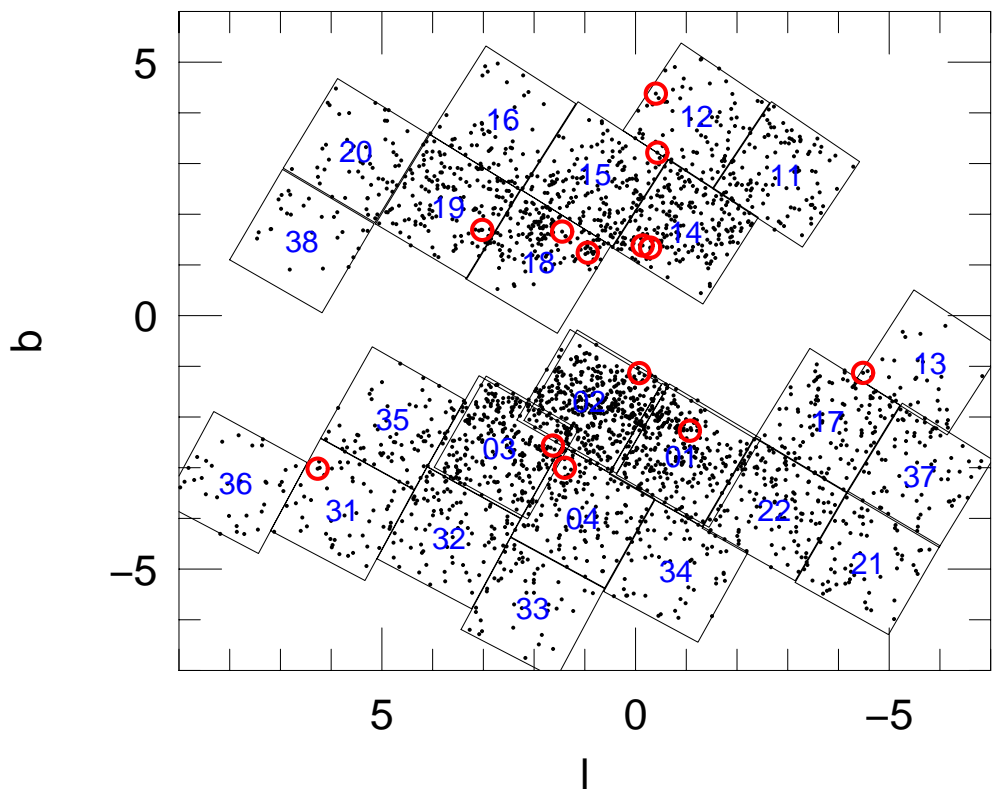}
\caption{The distribution of 2019 KMT FSPL events (red) in Galactic coordinates
compared to corresponding distribution of EventFinder events (black).
The FSPL events may appear to ``avoid'' areas of high event concentration.
However, while Figure~\ref{fig:ks} confirms this impression, it also
shows that this feature is not statistically significant.  
The black squares
outline the KMT fields, which are labeled with blue field numbers.
Note that to avoid clutter, fields BLG41, BLG42, and BLG43 are shown, 
but not labeled.  As shown, they lie toward slightly higher $b$ and
lower $l$ compared to the corresponding fields BLG01, BLG02, and BLG03.}
\label{fig:lb}
\end{figure}

\begin{figure}
\plotone{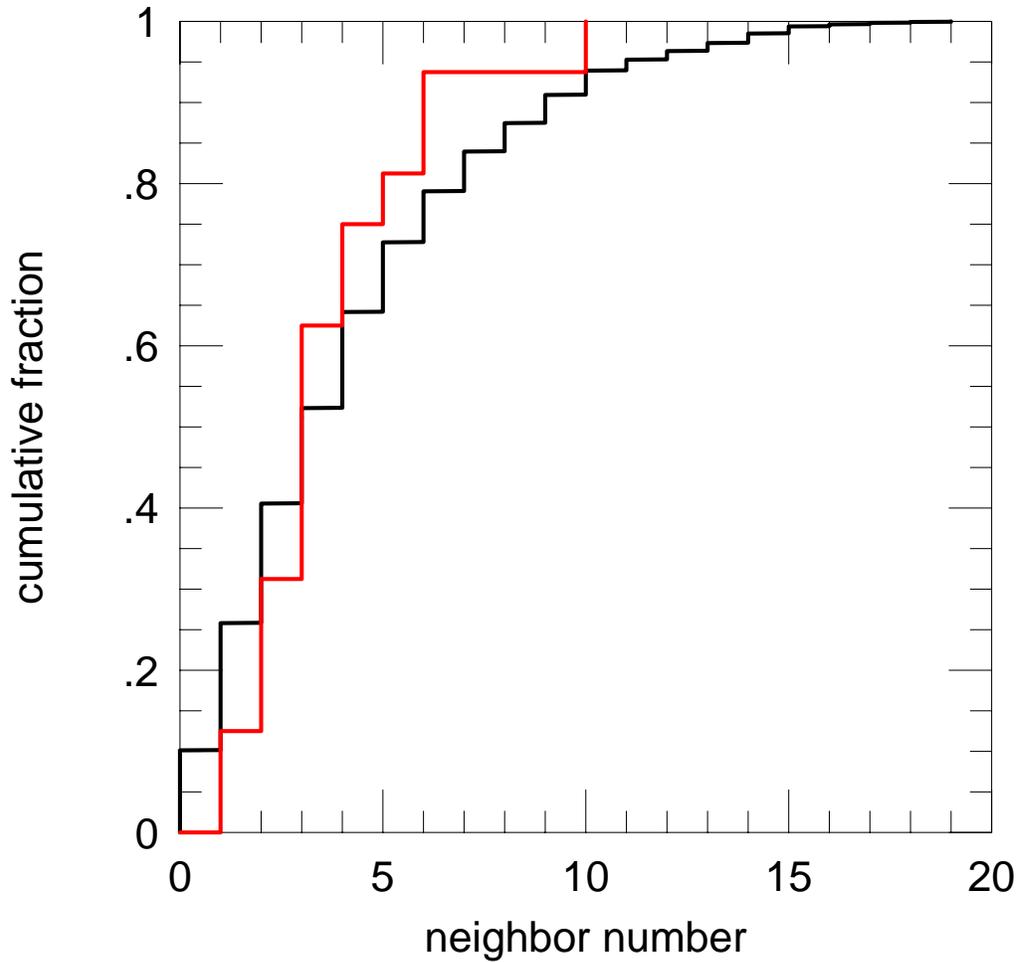}
\caption{Cumulative distributions of 2019 FSPL events (red) compared
to 2019 EventFinder events (black), ranked by number of EventFinder
neighbors within $10^\prime$ (roughly the size of the red circles
in Figure~\ref{fig:lb}).  While
there is a higher fraction of higher-neighbor-number 2019 EventFinder events
(in accord with the visual impression from Figure~\ref{fig:lb}),
this is not statistically significant.}
\label{fig:ks}
\end{figure}

Figure~\ref{fig:lb} shows the distribution of 2019 FSPL events in 
Galactic coordinates compared to 2019 EventFinder events.  By eye,
the FSPL events appear perhaps to avoid high-concentration areas of the
EventFinder distribution.  Figure~\ref{fig:ks} confirms that a larger
fraction of EventFinder events have large numbers of near neighbors,
but at the same time shows that this effect is not statistically
significant.

Figure~\ref{fig:lb} also shows that there are more FSPL events in 
the northern bulge (7) than the southern bulge (6), despite the
fact that only about 25\% of KMT observations are toward the northern
bulge.  This is somewhat surprising but is also not statistically
significant.  First, we expect that detection of finite-source
effects in giant-star events should be substantially less frequent in
fields with nominal cadences $\Gamma\geq 1\,{\rm hr}^{-1}$ than
those with $\Gamma\leq 0.4\,{\rm hr}^{-1}$, because the former
are well sampled over the peak while the latter are relatively
poorly sampled, given that typical giant-star $t_*\sim 7\,$hr.
And indeed, there are only three of 13 events in the latter category
(in fields BLG12, BLG13, and BLG31).  All of the remaining 10 lie
in the eight $\Gamma\geq 1\,{\rm hr}^{-1}$ fields that are grouped
closest to the Galactic center, i.e., BLG01/41, BLG02/42, BLG03/43 and
BLG04 in the south and BLG14, BLG15, BLG18, and BLG19 in the north.
There are four events in the first group and six in the second, which
is not significantly different.  
Thus there is no evidence for unexplained structure
in the on-sky distribution of FSPL events.

Tables~\ref{tab:selected} and \ref{tab:fspl} show two different estimates
of the dereddened source magnitude $I_{s,0}$.  The first,
$I_{s,0,\rm web} = I_{s,\rm web} - A_{I,\rm gonzalez}$ is derived from the 
pipeline-PSPL-fit
source flux and the cataloged extinction estimate derived from
\citet{gonzalez12}.  The second, from the CMD analysis,
$I_{s,0,\rm cmd} = (I_s - I_{\rm cl})_{\rm dophot} + I_{\rm cl,0,nataf}$,
is derived by adding the offset of the source relative to the clump
in a DoPhot-based CMD to the dereddened magnitude of the clump from
Table~1 of \citet{nataf13}.  Comparison shows that $I_{s,0,\rm web}$ is
systematically fainter than $I_{s,0,\rm cmd}$, and that the scatter
in these offsets is significantly larger than the absolute error
in the $I_{s,0,\rm cmd}$ estimate, which is typically 
$\sigma(I_{s,0,\rm cmd})\la 0.07\,$mag.  That is, the statistical
properties of the $I_{s,0}$ estimates used for event selection
are significantly different than the true values.  In order
to understand the potential impact of this difference, it is 
first necessary to identify its origin.

\begin{figure}
\plotone{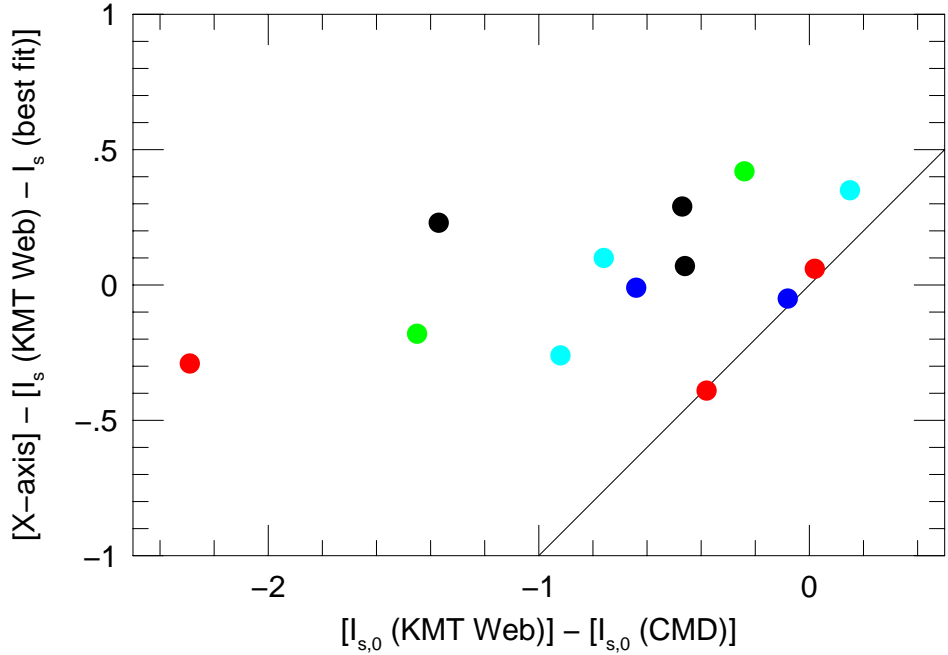}
\caption{The abscissa ``X-axis'' shows the difference between the 
estimates of the dereddened source  magnitude $I_{s,0}$ from
the candidate selection process (KMT Web) and from 
the final (CMD) analysis.  The ordinate shows the component of this
difference that is due to everything except the difference between
the pipeline-PSPL modeling of the event (without finite-source effects)
and the human-supervised modeling with finite-source effects.
The X-axis distribution displays large scatter and is strongly asymmetric.  
The Y-axis distribution has much smaller scatter and is symmetric about
zero.  This shows that essentially all of the systematic offset 
and most of the scatter is due to
the pipeline-PSPL-modeling error, $Z=X-Y$.  The values of $Z$ can be judged
for individual points by noting their horizontal distance to the 
black diagonal line.}
\label{fig:iscomp}
\end{figure}

With this aim, we define
\begin{equation}
X\equiv I_{s,0,\rm cmd} - I_{s,0,\rm web};
\qquad
Z\equiv I_{s,\rm best-fit-pys} - I_{s,\rm pipeline-pys} 
\qquad
Y\equiv X - Z.
\label{eqn:xy}
\end{equation}
That is, $X$ is the difference in $I_{s,0}$ estimates, $Z$ is the portion
of this difference that is due to a wrong pipeline-PSPL model, 
and $Y$ is the portion
that is due to everything else.  We will examine this ``everything else''
in detail below.  But first we note that Figure~\ref{fig:iscomp}, shows
that ``everything else'' has essentially zero mean offset
$\langle Y\rangle = 0.03\pm 0.07$ and relatively small scatter,
$\sigma(Y) = 0.25$.  This means that most of the scatter and all of the
systematic offset of $X$ in Figure~\ref{fig:iscomp} is due to incorrect
pipeline-PSPL modeling of the light curve.  
We will show that most of the scatter
from ``everything else'' is due to the extinction estimate and the
approximate calibration of KMT online photometry, both of which
impact $I_{s,0,\rm web};$ rather than $I_{s,0,\rm cmd}$.  

\begin{figure}
\plotone{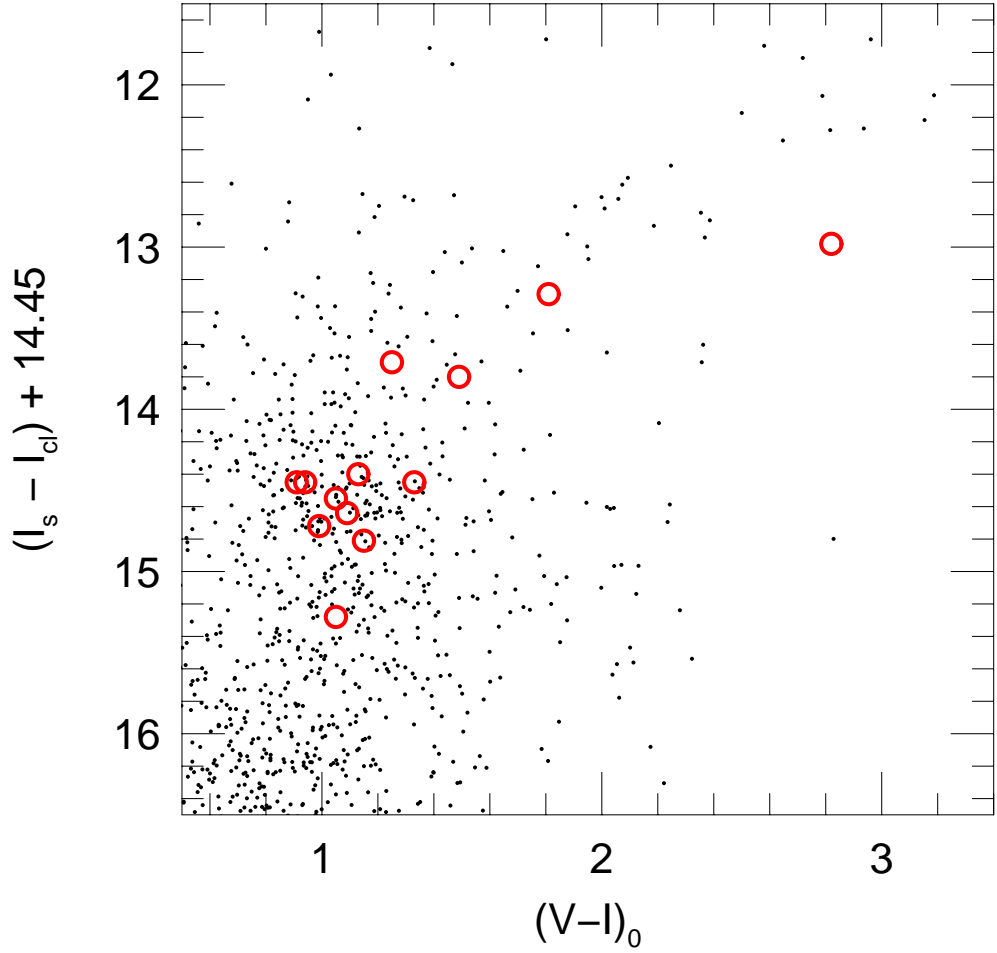}
\caption{Location of each of the 13 FSPL events relative to the
clump projected onto the dereddened CMD of KMT-2019-BLG-2555.
As expected, the FSPL events generally trace the giant branch and red
clump, but weighted toward brighter (so bigger, i.e., higher cross section)
stars.}
\label{fig:cmdall}
\end{figure}

Therefore, the main
consequence of the broad and asymmetric distribution of $X$ is that
it affects selection.  In particular, it removes of order a third of
%bbb: $16 > I_{s,0,\rm true} >15$ -> $16 > I_{s,0,\rm true} >15$
all candidates $16 > I_{s,0,\rm true} >15$, as well as a few that are brighter,
via Criterion (1)\footnote{As analyzed in the discussion of 
Table~\ref{tab:fspl},
all the events obey $\mu_\rel \la \mu_{\rm thresh}$, so the
automated selection of candidates for review is not seriously impacted
by inaccurate input parameters in this respect.  In any case,
as shown in that discussion, the $\mu_{\rm thresh}$ boundary is set
very conservatively, so that even this rare error would result in
deselection of very few viable candidates.
}.
The unwanted rejection of potential candidates will, of course,
adversely affect the size of the sample, but it will not in itself
affect its statistical character.  Each giant, regardless of how
it is selected, should be equally sensitive to isolated lenses,
regardless of their mass.  The exception would be if the pipeline-PSPL modeling
errors were more severe for low-mass events, so that more were
artificially driven over the selection boundary by this effect.
We see no evidence of this in our very small sample of two FFP candidates.
But even if there were such an effect, its overall impact would be
%bbb: $16 > I_{s,0,\rm true} >15$ -> $16 > I_{s,0,\rm true} >15$
small because the affected region of the CMD $16 > I_{s,0,\rm true} >15$
generates relatively few FSPL events, as we discuss in relation
to Figure~\ref{fig:cmdall}, below.  And, as just mentioned, only
about 1/3 of these are inadvertently eliminated.

We now turn to a more detailed investigation of the various independent
terms grouped under $Y$, i.e.,  ``everything else''.  Rearranging terms
in the equation $Y=X-Z$, we obtain
\begin{equation}
%[I_{s,0,\rm cmd} - I_{s,0,\rm web}] - I_{s,\rm best-fit-pys} - I_{s,\rm machine-pys} =
%aaa: added period marks at the ends of equation (12)
Y = [I_{s,\rm machine-pys} - I_{s,0,\rm web}] - [I_{s,\rm best-fit-pys} - I_{s,0,\rm cmd}].
\label{eqn:deltaIs}
\end{equation}
%A_{I,\rm gonzalez} - [I_{s,\rm best-fit-pys}] - I_{s,\rm best-fit-dop}] - I_{\rm cl,dop}] + I_{\rm cl,0,nat}
%A_{I,\rm gonzalez} - [K_{\rm pys} - K_{\rm dop} + \delta_{\rm fit}] - [I_{\rm cl,dop}] - I_{\rm cl,0,nat}] 
%A_{I,\rm gonzalez} - \delta_{\rm fit} - [I_{\rm cl,pys}] - I_{\rm cl,0,nat}] 
%A_{I,\rm gonzalez} - \delta_{\rm fit} - [I_{\rm cl,pys}] - I_{\rm cl,0,nat}] 
%A_{I,\rm gonzalez} - \delta_{\rm fit} - [I_{\rm cl,pys}] - I_{\rm cl}] - [I_{\rm cl} - I_{\rm cl,0,nat}] 
%A_{I,\rm gonzalez} - \delta_{\rm fit} - [K_{\rm pys}- K_{\rm true}] - [I_{\rm cl} - I_{\rm cl,0,nat}] 
%A_{I,\rm gonzalez} - \delta_{\rm fit} - \delta_{\rm K-pys} - [I_{\rm cl,0} - I_{\rm cl,0,nat}] 
%[A_{I,\rm gonzalez} -  A_{I,\rm true}] - \delta_{\rm fit} - \delta_{\rm K-pys} -[I_{\rm cl,0} - I_{\rm cl,0,nat}]
Then, after several substitutions, this can be evaluated as,
\begin{equation}
%aaa: added period marks at the ends of equation (13)
Y = [A_{I,\rm gonzalez} -  A_{I,\rm true}] - [{\rm Zpt}_{\rm pysis}-{\rm Zpt}_{\rm true}] 
+ \delta_{\rm centroid} + [I_{\rm cl,0,nataf} - I_{\rm cl,0,true}] 
+ \delta_{\rm pys-dop} .
\label{eqn:deltaIs2}
\end{equation}

These five terms are the error in $A_{I,\rm gonzalez}$ relative to the
true value, the error in the adopted KMT pySIS zero point ($I=28$)
relative to the true value, the error in fitting the clump centroid in
the CMD, the error in that $I_{\rm cl,0}$ relative to the true value,
and the offset between DoPhot and pySIS source-flux fit values
relative to the true value (established by, e.g., field star
comparison).  The last is typically $<0.01$ mag and can be ignored.
Apart from some unknown systematic offset, the penultimate term is
also small because the intrinsic variation over the bar is smooth and
the values in \citet{nataf13} Table~1 are established by averaging
many measurements.  The third term varies depending on the
density of the clump but is typically 0.05 mag.  The second term has
two principal components:  (1) the source counts for a star of
fixed magnitude vary smoothly over the KMT field, with an effective dispersion
of 0.05 mag, due to the optics, and (2) the transparency of the images
chosen for the template of a given event can vary.  We estimate the
total dispersion of this term as 0.07 mag.  Before examining the 
first term in detail, we note that given the total observed dispersion,
$\sigma(Y) = 0.25\,$mag, and our estimates of the other four terms, this
leaves a dispersion of 0.23 mag for the first term.  Several effects 
contribute.  The first is the measurement error in the underlying
\citet{gonzalez12} catalog, which includes errors in estimating the
mean $A_K$ over the ($2^\prime\times 2^\prime$) grid point that is
cataloged in the KMT database.  The second
is the difference between this true mean value and the actual value
at the location of the event due both to smooth variation of $A_K$
and patchy extinction.  The third is the difference between
$A_I$ and our adopted universal estimate of this as $A_I=7 A_K$.  These
effects can very plausibly account for the $0.2\,$mag ``observed
dispersion'' that was estimated above\footnote{The interactive
site http://mill.astro.puc.cl/BEAM/calculator.php does not quote
errors for $A_K$, but typically quotes $\sigma[E(J-K)\ga 0.1]$.
By our empirical estimate, $\sigma(A_K)\sim 0.03$ is several times
smaller.}.  In principle, each of the first, second, and fourth terms
could contain a systematic offset.  However, if such systematic
offsets are present, they happily cancel to within $\sim 0.06\,$mag.

The points in Figure~\ref{fig:iscomp} are color-coded according to
$z_0\equiv u_0/\rho$.  One may generally expect that the pipeline-PSPL fitting
program will be ``confused'' by strong finite source effects ($z\la 0.5$)
because it does not contain $\rho$ as a fitting parameter.  Indeed
the three most severe outliers at the left all lie in this regime:
KMT-2019-BLG-2555 ($z=0.334$), %0738
OGLE-2019-BLG-0953 ($z=0.479$), and %1221
KMT-2019-BLG-1143 ($z=0.077$). %0743 
However, there are many other events with similar $z_0$ that suffer
much smaller (or essentially no) pipeline-PSPL modeling problems.  With
one exception, we are not able to trace additional factors that
distinguish the outliers from the others.

The exception is KMT-2019-BLG-2555.  EventFinder assigned a catalog 
star to this event that was 4 mag fainter than the nearest catalog
star (and actual source) of the event, which confused the pipeline-PSPL fit.

Figure~\ref{fig:cmdall} shows the positions of all the FSPL source stars 
relative
to the clump centroid compared to the dereddened CMD of KMT-2019-BLG-2555.
This event was chosen for the comparison due to its relatively densely
%aaa: although it is does suffer relatively... -> although it does suffer relatively... 
populated CMD, although it does suffer relatively high extinction,
$A_I=3.71$.  Of course, this dereddening applies only to stars that
lie behind the full column of dust seen toward the clump, but these
account for the overwhelming majority of the field stars that lie
%bbb: ... of the CMD that is displayed. -> ... of the CMD that is NOT displayed.
in the part of the CMD that is displayed.  Figure~\ref{fig:cmdall} shows 
that 12 of the 13 FSPL sources closely follow the track of red giants
and clump giant stars from the CMD.  Eight of these 12 are tightly grouped
in the clump, which displays a strong overdensity in the field
star distribution.  One lies on the lower giant branch, below
the clump:  KMT-2019-BLG-0313.  This event
is projected close to the edge of a small dark cloud, so its position on the CMD
was estimated using a special procedure.  See Section~\ref{sec:kb190313}.
And three of the 12 lie on the upper giant branch.
%aaa: ... consistent with expectations, -> ... consistent with expectations.
This distribution is qualitatively consistent with expectations.
The clump stars
are somewhat more numerous than the lower giant branch stars, and they
are further favored both by their higher cross section and the fact that
(based on the analysis of Figure~\ref{fig:iscomp}) we expect the
selection process to eliminate about a third of the FSPL events
%bbb: $16 > I_{s,0,\rm true} >15$ -> $16 > I_{s,0,\rm true} >15$
with $16>I_{s,0,\rm true}>15$.
Similarly, the upper giant branch is substantially more thinly
populated than the clump but is favored by higher cross section.
There is also one ``outlier'': KMT-2019-BLG-1143 at the extreme right.
It lies about 0.75 mag below the roughly horizontal track of upper giant
branch field stars.  This could be explained either by the source being
exceptionally metal rich or by it being $\sim 3\,\kpc$ behind the mean
distance to the Galactic bar, either in the far disk or in the distant
part of the bar itself.
We further remark on this event in the notes on individual events,
Section~\ref{sec:kb191143}.

\subsubsection{{Distribution of $\theta_\e$}
\label{sec:distthetae}}

\begin{figure}
%ccc: path change
%\plotone{../eventfinder/thstar.eps}
\plotone{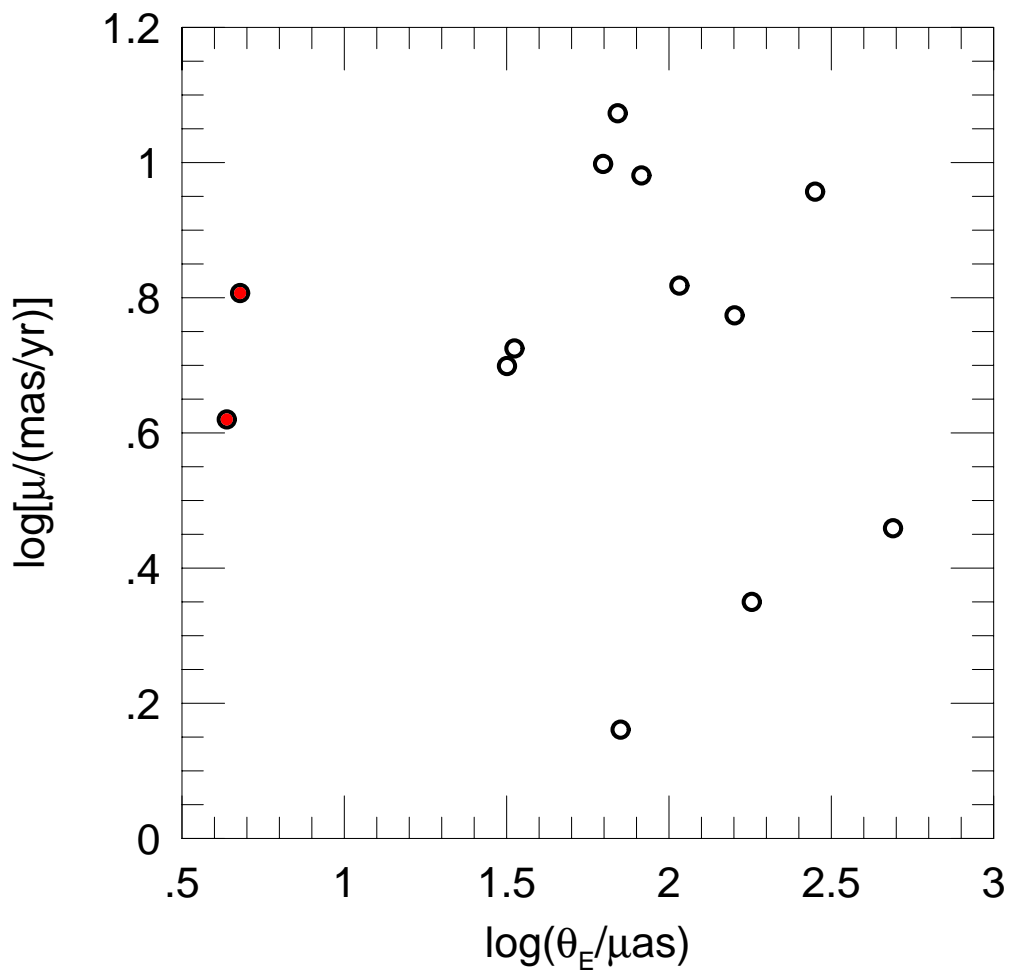}
\caption{Plot of lens-source relative
proper motion $\mu$ versus 
Einstein radius $\theta_\e$ 
for 13 FSPL giant source events from 2019.  The
median proper motion is $\mu_{\rm med} = 5.9\,\masyr$, which is typical
of microlensing events with measured $\mu$.  The two FFP candidates (red)
have proper motions straddling the median, i.e., 4.2 and 6.4 $\masyr$.
The two FFP candidates lie well separated from the rest of the giant-star
FSPL event in $\theta_\e$, which are otherwise evenly distributed
in $\log\theta_\e$.
}
\label{fig:thstar}
\end{figure}

Figure~\ref{fig:thstar} is a scatter plot 
of Einstein radius $\theta_\e$ versus lens-source relative
proper motion $\mu$.  The
median proper motion is $\mu_{\rm med} = 5.9\,\masyr$, which is typical
of microlensing events with measured $\mu$.  Leaving aside the two
events at the left edge of the distribution, the remaining 11 events
have Einstein radii in the range $31 \la \theta_\e/\muas \la 490$.
That is, if all had the same $\pi_\rel$,
then these lenses would span a factor $(490/31)^2 = 250$ in mass.
For example, for $\pi_\rel = 16\,\muas$, this mass range would
be: $7.7\,M_{\rm jup}$ to $1.8\,M_\odot$.  Of course, not all the
lenses have the same $\pi_\rel$, but this simple calculation
suggests that these FSPL events span a wide range of stellar and
brown-dwarf masses.  The two low-$\theta_\e$ events are FFP candidates.

\subsection{{Notes on Individual Events}
\label{sec:notes}}

\subsubsection{OGLE-2019-BLG-1182}
\label{sec:ob191182}

There are {\it Spitzer} data for this event, so it may ultimately yield
an isolated-object mass measurement.  However, the {\it Spitzer} data
begin 3.53 days after $t_{0,\oplus}$, i.e., at $u_{\oplus}=1.13$ for this
short $t_\e = 3.13\,$day event.  Hence, it is possible that the {\it Spitzer}
parallax will only yield a relatively large circular-arc constraint in the
$\bpi_\e$ plane \citep{gould19}.
%99.36 - 95.83 = 3.53 /3.13 = 1.13

\subsubsection{KMT-2019-BLG-2800}
\label{sec:kb192800}

%bbb: There are no magnified $V$-band points -> There is no magnified $V$-band point 
%ccc: according to the table3, t_E=2.146day rather than t_E=3.11day
There are no magnified $V$-band points for this short ($t_\e=2.14\,$day) event.
However, the fit to DoPhot \citep{dophot} reductions shows that it is
consistent with zero blending, so that the color can be estimated
from the baseline object.  We identify the baseline object on
a $[Z-K,K]$ VVV \citep{vvvcat} CMD, from which we determine that
it lies $\Delta(Z-K)=0.18$ mag redward of clump.  
Using an $(IZK)$ color-color diagram, we determine that this
corresponds to $\Delta(I-K)=0.23$, and then using \citet{bb88}, that
this implies $\Delta(V-I)=0.27$.  The resulting $\theta_*=8.1\,\muas$ leads
to a relatively high lens-source relative proper motion, 
$\mu_\rel = 11.8\,\masyr$.  However, this is less surprising after considering
that the Gaia baseline-object proper motion is 
$\bmu_{\rm base}(N,E) = (-10.28,-4.72)\pm (0.32,0.39)\,\masyr$.  
Given the low/zero blending, this measurement can be taken as a proxy
for $\bmu_s$.

\subsubsection{KMT-2019-BLG-0313}
\label{sec:kb190313}

In the finding chart, this event is seen to lie near the edge
of a small dark cloud, perhaps one of a string of such clouds that
extends south-west to north-east, diagonally through the field.
The cloud is far too small to form a CMD of stars of similar extinction,
which is the normal basis of the \citet{ob03262} technique.  Instead,
we form such a CMD from the larger field and then project the source
along the reddening vector, using a slope $R_{VI} = d I/d(V-I) = 1.37$
until it hits the lower giant branch at $[(V-I),M_I] = (1.05,0.71)$.

\subsubsection{KMT-2019-BLG-2528}
\label{sec:kb192528}

This event is almost entirely contained in ``pre-season'' data taken
at the end of the night, only from KMTC and only in $I$-band.  This
observing program was motivated to constrain the parallax measurement of
KMT-2018-BLG-1292 \citep{kb181292}, but was carried out in all western
KMTNet fields.  Hence, the source color cannot be measured from the
light curve.

Unfortunately, the fit to DoPhot \citep{dophot} reductions shows that
the source is blended, so we cannot simply derive the color from that
of the baseline object.  However, it is still the case that 
$\eta=f_b/f_{\rm base}= 0.166$ is relatively small.  We therefore begin
with a modified version of this approach (see Section~\ref{sec:kb192800}).

First, we find that the baseline object lies $\Delta(Z-K) = 0.49$
redward of the clump.  If we assume that the blend has the
same color as the baseline object (which is extremely red) then
we obtain from the $IZK$ color-color diagram that $\Delta(I-K) = 0.62$.
Then, following the same procedures as above, we derive 
$(V-I)_{\rm base,0} = 1.69$, and so $(V-I)_{s,0} = 1.69$. 

However, because the blend is 1.75 mag fainter in $I$ than the source,
hence $I_0\sim 14.88$, it most likely is a clump star or a first ascent
giant just below the clump, and hence would have $(V-I)_{0,b}\sim 1.04$
and thus $\Delta(I-K)_{0,b} = -0.03$ relative to the clump.  Then,
instead of the blend accounting for 16.6\% of the $K$ band
light of the baseline object, it would account for only 9.9\%.
Thus, the source would be redder yet by $\delta(I-K) = 0.08$ mag,
implying $(V-I)_{s,0} = 1.81$.  Finally, conceivably, the blended light
could be due to an extreme foreground object (e.g., the lens) in which
case it would account for a tiny fraction of the $K$-band flux from the
baseline object, and so $(V-I)_{s,0} \sim 1.9$.  We finally adopt
$(V-I)_{s,0} = 1.81$, recognizing that there is some additional uncertainty
in the color for this event.  However, this added uncertainty has no
material impact on the scientific conclusions in the current context.

\subsubsection{KMT-2019-BLG-1143}
\label{sec:kb191143}

This event has a complex discovery history and also posed some challenges
in measuring the source color.  The event designation KMT-2019-BLG-1143
derives from an alert posted to the KMT web page on 6 June 
(HJD$^\prime=8640.66$) as ``probable'' microlensing, 
when the event was at magnification $A=14.6$,
corresponding to an $I = 14.4$ ``difference star''.  This is surprisingly
late.  Moreover, the subsequent DIA light curve used to
classify alerts did not seem to confirm the microlensing interpretation,
and it was reclassified as ``not-ulens''.  Then the same event
was apparently ``rediscovered'' by EventFinder, with the DIA light curve 
tracing a very well defined and obvious microlensing event.  

This puzzling
discrepancy was resolved as follows.  The EventFinder catalog star lies
$3.5^{\prime\prime}$ roughly north of the AlertFinder catalog star.  The
AlertFinder program actually triggered on the former on 26 March, i.e.,
72 days earlier, when the $(A-1)=0.22$ magnification produced a difference
star of $I=18.9$.  The candidate was then ``misclassified'' as
a variable\footnote{It is actually a low-level variable, but the microlensing
signal already substantially exceeded the level of source variability.}, 
and was thus
not shown to the operator again as it further evolved.  Then, as the
event neared peak, it became so bright that the excess flux inside
the more southerly catalog-star aperture rose sufficiently to trigger
a human review.  Finally, after the ``rediscovery'', the program that
cross matches AlertFinder and EventFinder discoveries identified them
as the same event.  Note that even if the KMT-2019-BLG-1143 had not
been ``rediscovered'' by EventFinder, it would have been recognized
as having the wrong coordinate centroid, which would have been 
corrected, in the end-of-year re-reductions.  Unfortunately, in 2019
there was not sufficient computing power to properly re-centroid
already discovered events contemporaneously with other real-time tasks.
Otherwise, the event would have almost certainly been chosen as
a {\it Spitzer} target.

The DoPhot fit shows that the source is blended with another much-bluer star
that is $\Delta I= 1.36\pm 0.07$ mag fainter.  The blended star is
well localized to lie in the clump, but the source $V$-band flux
is not reliably measured.  We therefore adopt a similar approach as
for KMT-2019-BLG-2528 (Section~\ref{sec:kb192528}), with two adjustments.
First, we assume that the blend is a clump giant rather than considering
a range of possibilities.  Second, we adopt  the 2MASS measurement
$K=10.375$ in place of the VVV measurement $K=10.963$ because the latter is
saturated.  We then find that the source 
lies $\Delta(Z-K)=1.07$ mag redward of clump, which we transform to
to $\Delta(I-K)=1.40$, and finally $\Delta(V-I)=1.76$.  Because the
source is an M giant, we use the M-giant color/surface-brightness relation
of \citet{groenewegen04} to evaluate $\theta_*$.

We note that KMT-2019-BLG-1143, which has the largest Einstein
radius of the sample, $\theta_\e=489\,\muas$, also has a well-measured
annual microlensing parallax, $\bpi_\e(N,E) = (0.146,0.278)$.  
Together these yield a lens mass $M=\theta_\e/\kappa\pi_\e=0.19\,M_\odot$
and relative parallax $\pi_\rel = 0.154\,\mas$.  Assuming that the
source is in the bulge at $\pi_s = 0.12\,\mas$, the lens distance is then
$D_L = 3.7\,\kpc.$

\section{{Discussion}
\label{sec:discuss}}

\subsection{{Fourth FSPL FFP Candidate with $\theta_\e<10\,\muas$}
\label{sec:4thFFP}}

KMT-2019-BLG-2073 is the fourth FFP candidate with measured
$\theta_\e<10\,\muas$.  The previous three had 3-4 year
intervals between them:
OGLE-2012-BLG-1323 \citep{ob121323}, OGLE-2016-BLG-1540 \citep{ob161540},
and OGLE-2019-BLG-0551 \citep{ob190551}.  The discovery of
KMT-2019-BLG-2073 in the same year as OGLE-2019-BLG-0551 \citep{ob190551}
may be just due to chance, but it could also reflect improved
sensitivity of the KMT selection procedures to short FSPL events
(e.g., searches for events with $t_\eff<1\,$day).
KMT-2019-BLG-2073 was found independently by the AlertFinder in real time,
and by the EventFinder in post-season analysis.  The AlertFinder
was in full operation for the first time in 2019.  (In 2018, it
essentially operated only in the northern bulge and only for about
half the season).  The EventFinder found it as a $t_\eff = 0.42\,$day
event.  In 2016, the search was conducted only for $t_\eff \geq 1\,$day,
and this remained so for 60\% of sources in 2017.  Hence, it is 
possible that FFP candidate events remain undiscovered in the database of KMT 
light curves.

\subsection{{Future Adaptive Optics Search for Host}
\label{sec:hostao}}

A key open question for KMT-2019-BLG-2073 is whether or not the lens is indeed a free-floating planet rather than a planet on a wide orbit. In Section \ref{sec:host}, we discussed the limits on a possible host star that can be inferred from the microlensing light curve alone. A more powerful test is to image the system directly to search for light from a host star. Due to the brightness of the source star, this is \HL{only really feasible} by waiting for the lens and source to separate, so \HL{that} light from the lens can be resolved.

\HL{Because the source is a bright giant $(J,H,K)_s=(15.28,14.16,13.72)$,
a high contrast-ratio sensitivity 
is required to detect even the brightest putative host, i.e. a main-sequence star in the Galactic bulge.  For example, for a Sun-like host in the bulge,
%A (I,J,H,K) =    (3.77,1.56,0.83,0.54)  (l,b)=(-0.07,-1.13)
%M_(I,J,H,K)_sun =(4.15,3.66,3.32,3.27)
%A_I + M_I       = (7.92,5.22,4.15,3.81)
the extincted apparent magnitudes would be,
$(J,H,K)_{\rm sun-like}\simeq  (19.7,18.7,18.3)$, and it would therefore
have contrast ratios $\Delta(J,H,K) = (4.4,4.5,4.6)$ magnitudes.}

\HL{As a point of comparison, \citet{bowler15} achieved contrast
ratios of $\Delta K=5$ magnitudes for several of their targets
at separations of $\Delta\theta = 100\,$mas using a coronagraph
on Keck.  This comparison immediately raises three issues.
First, at its measured lens-source relative proper motion
$\mu_{\rm rel} = 6.4\,\masyr$, the system will not reach 100 mas separation
until 2035.  Second, at $K_s= 13.7$, the source star is too faint
for the Keck coronagraph.  While conventional adaptive optics (AO)
observations would be possible, their contrast performance is inferior
to that of coronagraphs.  Third, \citet{bowler15} found that
the next significant improvement in contrast ratio came at 
$\Delta\theta=300\,$mas, where it increased to about $\Delta K=11$
magnitudes.  See their Figure~3.}

\HL{This comparison makes it clear that very little can be accomplished
with current instruments for many decades.  Long before this time,
next-generation extremely large telescopes (ELTs) will come on line.
Because planet imaging is a major driver for these projects, it is
likely that they will mount coronagraphs.  To be specific, we will
consider that these are operational in 2032, i.e., when
$\Delta\theta= 83\,$mas.  Scaling to the experience of \citet{bowler15},
we estimate that the 39m European ELT (EELT) will achieve 
$\Delta K = 11$ mag of contrast at $\Delta\theta = 300/3.9 = 77\,$mas.
That is, it could probe for putative bulge hosts down to 
$K = 24.7$, i.e., $M_K=9.7$, which is below the bottom of the main
sequence.  While, as mentioned above, the source star is too faint
for the Keck coronagraph, the magnitude limit for an EELT coronagraph
should increase by $5\log(39/10)= 3.0\,$mag, even without further
improvements in instrument design.  Thus it is plausible that an
EELT coronagraph could probe for hosts down to the bottom of the
main sequence.}

\HL{We note that several other FFP candidates have favorable characteristics
for EELT observation, including OGLE-2012-BLG-1323, KMT-2017-BLG-2820,
OGLE-2016-BLG-1928, and 
OGLE-2016-BLG-1540, which will have separations of 
$\Delta\theta=112\,$mas, 128$\,$mas, 138$\,$mas, and 168$\,$mas, respectively.}

\subsection{{Future FFP Statistical Studies}
\label{sec:FutureFFP}}

In Section~\ref{sec:stats}, we have presented a detailed outline of
a statistical approach to measuring the relative frequency of
FFP's compared to stars and brown dwarfs. Our approach has two main features. First, it uses the $\theta_\e$ distribution, which is a more robust measure of lens mass than the $t_\e$ distribution. Second, we focus on events with giant sources, which are more likely to yield the finite source effects needed to measure $\theta_\e$ and have a number of other practical advantages (described in Section \ref{sec:stats}).

\begin{figure}
%ccc: path change
%\plotone{../eventfinder/cum.eps}
\plotone{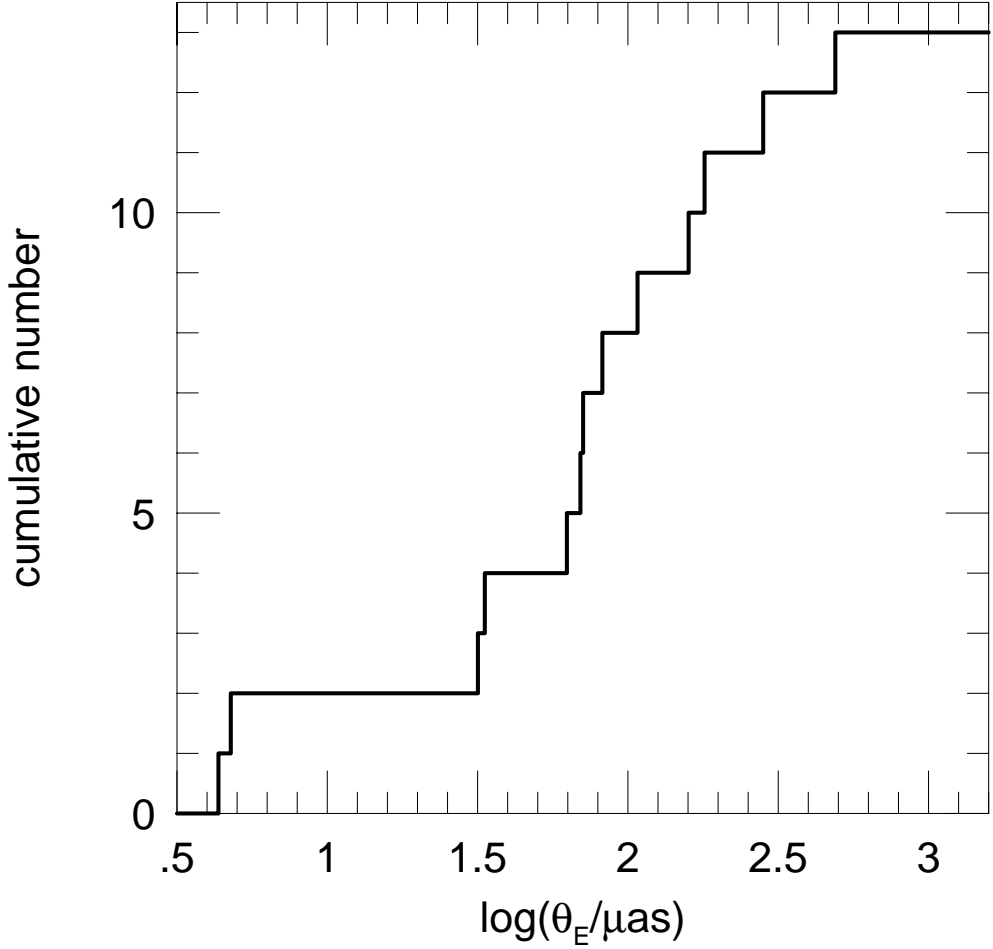}
\caption{Cumulative distribution of log Einstein radii $\log\theta_\e$
for 13 giant-star FSPL events from 2019.  The cumulative distribution is
consistent with linear for $\theta_\e\ga 30\,\muas$, i.e., a flat 
differential distribution.  Because 
$M = 0.0069\,M_\odot (\theta_\e/30\,\muas)^2/(\pi_\rel/16\,\muas)$ this
uniform-in-log distribution corresponds to a wide range of
stellar and brown dwarf masses. The sudden plateau below 
$\theta_\e< 30\,\muas$, may reflect a real absence of lower-mass
objects.  However, the apparent excess events (within this ``desert'')
at $\theta_\e\sim 5\,\muas$
cannot be used to make inferences about FFPs, due to publication bias.
}
\label{fig:cum}
\end{figure}

We apply this method to the 2019 season as a case study to identify and explore any unexpected challenges in this approach, we 
%aaa: changed by professor, 42 giant -> 43 giant
derived a sample of 43 giant-source events from the KMT event
%aaa: ... database, which is ... -> ... database, which was ...
database, which was based on the union of events found by the
AlertFinder and the EventFinder, together with a supplemental
search using a version EventFinder that was tuned to giant source-star
events with finite-source effects.  We found that 13 were FSPL events,
implying that they yielded measurements of $\rho$ and so
$\theta_*$, $\theta_\e$, and $\mu$.  We found a factor six gap
in $\theta_\e$ between the two FFP candidates and the 11 other events.
The cumulative distribution of the latter (Figure~\ref{fig:cum})
is consistent with being linear in $\log\theta_\e$.  That is, 
$d N/d\log\theta_\e \sim\,$const.

We emphasized at the outset (and we repeat here) that no scientific
conclusions about FFPs can be drawn from this sample because
it was motivated by an apparently ``large number'' (i.e., 2) of FFPs
during the 2019 season, and so suffers from publication bias.
In addition, to draw statistical conclusions about FFPs, one would
have to study the selection function of low-$\theta_\e$ events in the
underlying KMT database.  As $\theta_\e$ is reduced, 
an FSPL light curve will become dominated by
finite source effects rather than the Paczy\'nski curve.  Thus, the
EventFinder and AlertFinder algorithms (built for PSPL curves) will
eventually fail to identify the events as potential microlensing.
And, even before that happens,
the operator may fail to correctly classify the event as microlensing.
Application of this approach to the full KMTNet sample and more detailed characterization of the selection function (e.g., through an injection-recovery test) will be the subject of future work, which is already underway \citep[e.g., ][]{kb172820}.

However, neither of these concerns weighs heavily for the sample
of larger $\theta_\e$ events.  Of these 11 events with FSPL
effects, all were found by the EventFinder, and all but three
(KMT-2019-BLG-2528, KMT-2019-BLG-2800,
and KMT-2019-BLG-2555) were found by the AlertFinder.
The first of these occurred in ``pre-season'' data, long before the
AlertFinder started 2019 operations.  The second peaked at 
HJD$^\prime = 8759$, i.e., Oct 2, and so 27 days after the AlertFinder had
ceased operations.  The third event peaked at HJD$^\prime = 8581$, 
i.e., Apr 7.  This was 11 days after the AlertFinder began 2019 operations,
so this event should have been found.  Nevertheless, the failure rate for
these events, which are typically both bright and relatively high-magnification
is low.  So the chance they would be missed by both search
algorithms, or lie in the fraction
of the season where they were only searched by EventFinder and missed,
is also low.  Furthermore all of the FSPL events were rediscovered
by the special supplemental search that we carried out, except for the
two that were excluded from this search because they had $t_\eff > 5\,$ days
(and so were expected to be easily detected in the regular searches).
The first concern (publication bias) also does not apply to this subsample
of 11: almost nothing was known about the subsample 
prior to undertaking this investigation.

We prefer to wait for a larger, multi-season sample of the higher-$\theta_\e$
events before undertaking a systematic investigation.  However, here
we would like to point out two robust features of the 2019 sample.
First the fact that the $\theta_\e$ distribution is approximately uniform
in $\log\theta_\e$ implies that lens mass distribution is consistent with being
uniform in $\log M$.  As we have emphasized, the event rate for FSPL events
is directly proportional to their frequency.  Hence, future statistical
studies on a larger sample could probe this mass function.

Second, the paucity of lenses with $\theta_\e\la 30\,\muas$ appears to be
real.  The first two events above this gap
(KMT-2019-BLG-0703 and KMT-2019-BLG-2555) are bright, 
relatively high-magnification
events that were easily selected for inspection by the
EventFinder algorithm, and easily recognizable as microlensing
in the completed-event visual inspection.  Events like
these are not likely to escape detection.  Hence,
this gap in the $\theta_\e$ distribution could reflect a dip
in the mass function of isolated objects. A larger statistical
sample will clarify this. Even better would be to directly measure
masses for FFPs from a combination of
$\theta_\e$ from finite source effects and $\pi_\e$ from satellite parallaxes \citep{gould21}. The experience gained now from FSPL searches is essential to planning such endeavors.

{\subsection{FFPs or Wide-Orbit Planets?}
\label{sec:wide-orbit}}

The main caveat with searches for microlensing FFPs is that, generally, one cannot be certain from the microlensing light curve alone that a planetary-mass lens is truly isolated. There are several ways to investigate whether or not the planetary-mass lens population is free-floating as opposed to having host stars at large separations \citep[][also discuss this issue in detail]{kb172820}. First, one can investigate the continuum of planets with increasing $s$ \citep[e.g., ][]{poleski18, poleski20} to see if this population is consistent with being an extension of the known population of bound planets. Second, the situation should be much clearer for the FFP candidate population probed by {\em Roman} \citep{johnson20}, because {\em Roman} will have FFP events with dwarf sources and will often have a sufficiently high-resolution to resolve out unrelated blends. Hence, {\em Roman} FFP candidates will either have detections of excess light due to a host star or much stronger limits on such light.

Third, the FFP candidates identified from ground-based FSPL events can be followed up using high-resolution imaging to search for host stars, as discussed in Section \ref{sec:hostao} for KMT-2019-BLG-2073. Because $\mu_{\rm rel}$ is directly measured for FSPL events, the separation between the lens and source can be predicted as a function of time. Hence, once enough time has passed, high-resolution imaging can be used to search for the flux from a host star. This ability to search for host stars is another major advantage of a $\theta_\e$-based study over one based on $t_\e$ (for which $\mu_{\rm rel}$ is unknown, so it can be difficult or impossible to definitively associate or exclude a given source with the event).

There are two caveats. First, the larger the separation of the lens and source, the more likely it is for an unrelated star to lie at the expected separation from the source. Second, if the planet is on a very wide (e.g., 1000 au) orbit, the host star will be sufficiently displaced from the lens to distort the measurement.

However, both of those caveats can be overcome simply by taking two epochs of high-resolution imaging. This will show that the magnitude of the proper motion matches that expected from the microlensing light curve. Furthermore, as discussed in \citet{kb172820} it can distinguish between planets in Kepler-like ($\sim 100$ au) and Oort-like ($\sim 1000$ au) orbits because while the magnitude of the host proper motion will be correct, the direction will be displaced from the direction of the source in proportion to the semi-major axis \citep[see also ][]{gould16}.

% As we saw for KMT-2019-BLG-2073, the light curve places significant, but not comprehensive, constraints on possible host stars.

\acknowledgments 
Work by AG was supported by JPL grant 1500811.
This research has made use of the KMTNet system operated by the Korea
Astronomy and Space Science Institute (KASI) and the data were obtained at
three host sites of CTIO in Chile, SAAO in South Africa, and SSO in
Australia.
%
%aaa: grant -> grants, added grant serial number 2019R1A2C2085965
Work by C.H. was supported by the grants of National Research Foundation 
of Korea (2017R1A4A1015178  and 2019R1A2C2085965). 

\begin{deluxetable}{lcc}
\tablecolumns{3} \tablewidth{0pc} \tablecaption{\textsc{Mean
parameters for 1L1S models}} \tablehead{ \colhead{Parameters} &
\colhead{1L1S} & \colhead{1L1S ($f_B$=0)} } \startdata
  $\chi^2/\rm{dof}$               &4548.957/4549         &4549.197/4549        \\
  $t_0$ $(\rm{HJD}^{\prime})$     &8708.598 $\pm$ 0.005  &8708.599 $\pm$ 0.004 \\
  $u_0$                           &0.241 $\pm$ 0.170     &0.163 $\pm$ 0.103    \\
  $t_{\rm E}$ $(\rm{days})$       &0.267 $\pm$ 0.026     &0.272 $\pm$ 0.007    \\
  $\rho$                          &1.184 $\pm$ 0.142     &1.138 $\pm$ 0.012    \\
  $f_S({\rm KMTC})$               &0.947 $\pm$ 0.196     &0.873 $\pm$ 0.001    \\
  $f_B({\rm KMTC})$               &-0.073 $\pm$ 0.196    &-                    \\
  $t_*$ $(\rm{days})$             &0.313 $\pm$ 0.011     &0.310 $\pm$ 0.005    \\
  $\hat{S}$                       &0.673 $\pm$ 0.035     &0.674 $\pm$ 0.014    \\
\enddata
\tablecomments{$t_*\equiv\rho t_\e$ and $\hat{S}\equiv
f_S/\rho^2$are derived quantities and are not fitted independently.
All fluxes are on an 18th magnitude scale, e.g., $I_S=
18-2.5\,\log(f_S)$.} \label{tab:single}
\end{deluxetable}

 \begin{deluxetable}{llrlrrrrrr}
 \tablecolumns{10} \tablewidth{0pc}
 \tablecaption{\textsc{Events selected for            investigation}}
 \tablehead{\colhead{KMT Name} & \colhead{Name} &
 \colhead{$\mu_{\rm thresh}$} & 
 \colhead{$\rho$ meas?} &
 \colhead{$u_0$} & \colhead{$t_\e$} &
 \colhead{$I_s$} & \colhead{$A_I$} &
 \colhead{$I_{s,0}$} & \colhead{$\Delta\chi^2$} }
 \startdata
KB191820 & OB191182 & 24.23 & Yes & 0.024 &   3.38 & 16.17 & 1.44 & 14.73 &  11187 \\
KB192800 & KB192800 & 19.13 & Yes & 0.048 &   2.22 & 16.69 & 2.04 & 14.65 &   9392 \\
KB191653 & KB191653 & 16.50 & Yes & 0.012 &   9.22 & 17.41 & 2.52 & 14.89 &   3690 \\
KB190313 & KB190313 & 15.52 & Yes & 0.032 &   2.65 & 19.97 & 4.37 & 15.60 &   5305 \\
KB190703 & KB190703 & 15.28 & Yes & 0.035 &   3.32 & 17.78 & 2.83 & 14.95 &  12738 \\
KB190853 & OB190726 & 12.66 & Yes & 0.013 &  11.72 & 17.05 & 2.28 & 14.77 &  13201 \\
KB192073 & KB192073 & 11.17 & Yes & 0.324 &   0.50 & 18.68 & 3.77 & 14.91 &   2663 \\
KB192528 & KB192528 &  8.40 & Yes & 0.022 &  14.55 & 17.31 & 3.26 & 14.05 &    973 \\
KB192555 & KB192555 &  6.89 & Yes & 0.014 &  11.35 & 19.71 & 3.71 & 16.00 &   2450 \\
KBS0111  & KBS0111  &  6.10 & No  & 0.088 &   2.96 & 21.52 & 6.33 & 15.19 &   1754 \\
KB190352 & KB190352 &  5.47 & No  & 0.044 &   8.13 & 17.96 & 3.22 & 14.74 &   4276 \\
KB192291 & OB191408 &  5.43 & No  & 0.109 &   2.03 & 17.86 & 2.06 & 15.80 &   1581 \\
KB191335 & KB191335 &  5.23 & No  & 0.024 &  14.01 & 17.65 & 2.68 & 14.97 &   6777 \\
KB191054 & KB191054 &  4.89 & No  & 0.017 &  16.28 & 19.78 & 4.24 & 15.54 &   4690 \\
KB190061 & OB190204 &  4.32 & No  & 0.053 &   5.62 & 16.97 & 1.32 & 15.65 &   7140 \\
KB192255 & OB191415 &  3.54 & No  & 0.166 &   6.35 & 15.00 & 1.66 & 13.34 &  17362 \\
KB190519 & OB190551 &  3.46 & Yes & 0.817 &   1.30 & 14.55 & 1.18 & 13.37 &   2180 \\
KB192368 & MB19006  &  3.05 & No  & 0.087 &  12.68 & 15.19 & 1.63 & 13.56 &  23787 \\
KB192542 & OB191459 &  2.87 & No  & 0.279 &   2.41 & 17.76 & 2.99 & 14.77 &   9479 \\
KB191315 & OB190953 &  2.86 & Yes & 0.005 &  78.29 & 19.36 & 3.41 & 15.95 &   2876 \\
KB191143 & KB191143 &  2.34 & Yes & 0.005 & 200.00 & 18.86 & 4.51 & 14.35 &  12675 \\
KB191053 & KB191053 &  2.29 & No  & 0.493 &   5.03 & 16.84 & 4.41 & 12.43 &  14007 \\
KB190289 & KB190289 &  2.24 & No  & 0.047 &  34.47 & 16.79 & 3.39 & 13.40 &   7111 \\
KB190141 & OB190171 &  1.82 & No  & 0.025 &  36.56 & 18.22 & 3.13 & 15.09 &  14683 \\
KB190863 & KB190863 &  1.81 & No  & 0.083 &  18.75 & 16.96 & 3.01 & 13.95 &  13343 \\
%aaa: added OB190382
KB192376 & OB190382 &  1.80 & No  & 0.388 &   2.94 & 16.22 & 1.58 & 14.64 &   2689 \\
KB191039 & OB190791 &  1.73 & No  & 0.366 &   2.00 & 17.84 & 2.15 & 15.69 &  25760 \\
KB190527 & KB190527 &  1.65 & Yes & 0.033 &  26.73 & 18.66 & 3.28 & 15.38 &   4147 \\
KB190007 & OB190140 &  1.63 & No  & 0.240 &   9.95 & 15.95 & 2.71 & 13.24 &  32049 \\
KB190732 & OB190698 &  1.62 & No  & 0.275 &   3.99 & 16.27 & 1.32 & 14.95 &  20609 \\
KB191030 & KB191030 &  1.54 & No  & 0.154 &   9.27 & 17.07 & 2.58 & 14.49 &  25535 \\
KB192074 & KB192074 &  1.49 & No  & 0.067 &  18.14 & 18.45 & 3.54 & 14.91 &  11579 \\
KB190125 & OB190268 &  1.42 & No  & 0.964 &   1.08 & 17.02 & 1.67 & 15.35 &   2568 \\
KB192422 & OB191462 &  1.30 & No  & 0.181 &   9.10 & 18.87 & 4.33 & 14.54 &  23921 \\
KB192297 & KB192297 &  1.27 & No  & 0.406 &   4.37 & 20.00 & 5.57 & 14.43 &   2810 \\
KB192115 & OB191307 &  1.25 & No  & 0.045 &  23.81 & 16.98 & 1.41 & 15.57 &   4363 \\
KB190424 & KB190424 &  1.21 & No  & 0.210 &   5.27 & 18.94 & 3.37 & 15.57 &   6724 \\
KB191477 & KB191477 &  1.12 & No  & 0.289 &   3.84 & 20.46 & 4.73 & 15.73 &    689 \\
KB190607 & KB190607 &  1.11 & No  & 0.057 &  30.01 & 16.63 & 1.83 & 14.80 &  11080 \\
KB191677 & KB191677 &  1.08 & No  & 0.127 &  11.65 & 20.66 & 5.49 & 15.17 &   4728 \\
KB190788 & KB190788 &  1.06 & No  & 0.209 &   7.90 & 19.03 & 4.04 & 14.99 &   8886 \\
KB192263 & OB191413 &  1.02 & No  & 0.821 &   4.76 & 14.63 & 1.43 & 13.20 & 114120 \\
KB192220 & KB192220 &  1.01 & No  & 0.783 &   3.85 & 19.99 & 6.22 & 13.77 &    642 \\
 \enddata
 \tablecomments{Event names are abbreviations for,
  e.g., KMT-2019-BLG-1820, OGLE-2019-BLG-1182, and 
  MOA-2019-BLG-006. \HL{In "KBS0111", "S" indicates this event was identified in the "Supplemental" search described in Section \ref{sec:supplement}.}}
 \label{tab:selected}
 \end{deluxetable}

 \begin{deluxetable}{llrrrrr}
 \tablecolumns{7} \tablewidth{0pc}
 \tablecaption{\textsc{Microlens Parameters for FSPL giant-star events}}
 \tablehead{\colhead{Name} & \colhead{KMT Name} &
 \colhead{$t_0$} & 
 \colhead{$u_0$} &
 \colhead{$t_\e$} & \colhead{$\rho$} &
 \colhead{$f_{s,\rm KMTC}$} }
 \startdata
OB191182 & KB191820 & 8695.82629 &  0.01998 &  3.129 & 0.07013 & 5.5886 \\
         & (errors) &    0.00045 &  0.00088 &  0.014 & 0.00046 & 0.0384 \\
%aaa: revised KB192800 parameter
% original file
% KB192800 & KB192800 & 8695.82726 &  0.02394 &  3.115 & 0.07144 & 5.0206 \\
%          & (errors) &    0.00042 &  0.00069 &  0.013 & 0.00042 & 0.0361 \\
KB192800 & KB192800 & 8758.89773 &  0.04534 &  2.146 & 0.11617 & 3.3195 \\
         & (errors) &    0.00364 &  0.01025 &  0.055 & 0.00352 & 0.1647 \\
KB191653 & KB191653 & 8684.67492 &  0.00232 &  5.975 & 0.06227 & 3.4576 \\
         & (errors) &    0.00097 &  0.00000 &  0.044 & 0.00056 & 0.0323 \\
KB190313 & KB190313 & 8578.79642 &  0.06479 &  2.018 & 0.08018 & 0.2269 \\
         & (errors) &    0.00083 &  0.00744 &  0.139 & 0.00699 & 0.0707 \\
KB190703 & KB190703 & 8613.40361 &  0.07759 &  2.312 & 0.17115 & 2.2435 \\
         & (errors) &    0.00088 &  0.00333 &  0.028 & 0.00240 & 0.0446 \\
OB190726 & KB190853 & 8639.12940 & -0.01649 & 11.315 & 0.02025 & 2.4698 \\
         & (errors) &    0.00032 &  0.00009 &  0.032 & 0.00011 & 0.0084 \\
KB192528 & KB192528 & 8525.27942 &  0.07814 &  9.788 & 0.12306 & 3.4655 \\
         & (errors) &    0.00544 &  0.00337 &  0.181 & 0.00239 & 0.0979 \\
KB192555 & KB192555 & 8581.05635 &  0.10038 &  2.302 & 0.30019 & 2.2663 \\
         & (errors) &    0.00160 &  0.01421 &  0.059 & 0.00969 & 0.1087 \\
OB190953 & KB191315 & 8662.12889 & -0.01594 & 29.253 & 0.03330 & 0.9185 \\
         & (errors) &    0.00191 &  0.00076 &  1.035 & 0.00123 & 0.0361 \\
KB191143 & KB191143 & 8644.90325 &  0.00668 & 62.192 & 0.06143 & 1.9834 \\
         & (errors) &    0.00145 &  0.00052 &  0.329 & 0.00035 & 0.0124 \\
KB190527 & KB190527 & 8620.78990 &  0.07092 & 17.955 & 0.07107 & 0.9385 \\
         & (errors) &    0.00343 &  0.00178 &  0.234 & 0.00246 & 0.0170 \\
 \enddata
 \tablecomments{Event names are abbreviations for,
  e.g., KMT-2019-BLG-1820.
   Fluxes are in units of $I=18$ 
 system.  For KB192073, see Table \ref{tab:single}.
 For OB190551 (=KB190519), see \citet{ob190551}.}
 \label{tab:parms}
 \end{deluxetable}

 \begin{deluxetable}{llrrrrrrrrr}
 \rotate
 \tablecolumns{11} \tablewidth{0pc}
 \tablecaption{\textsc{CMD and Derived Parameters for FSPL giant-star events}}
 \tablehead{\colhead{Name} & \colhead{KMT Name} &
 \colhead{RA} & \colhead{Dec} &
 \colhead{$(V-I)_0$} & 
 \colhead{$I_0$} & 
 \colhead{$\theta_*$} & 
 \colhead{$\theta_\e$} &
 \colhead{$\mu_\rel$} &                                          \colhead{$\mu_{\rm thresh}$} &
 \colhead{$z=u_0/\rho$} }
 \startdata
OB191182 & KB191820 & 18:00:41.45 & -29:15:42.80 & 1.15 & 14.75 & 5.86 &  82.19 &  9.58 & 24.23 & 0.285 \\
KB192800 & KB192800 & 18:11:24.09 & -25:01:07.25 & 1.33 & 14.27 & 8.08 &  69.54 & 11.84 & 19.13 & 0.390 \\
KB191653 & KB191653 & 17:32:12.87 & -27:35:16.91 & 1.13 & 14.42 & 6.58 & 107.60 &  6.58 & 16.50 & 0.037 \\
KB190313 & KB190313 & 17:42:38.85 & -26:50:16.33 & 1.05 & 15.22 & 4.30 &  62.68 &  9.95 & 15.52 & 0.808 \\
KB190703 & KB190703 & 17:46:11.29 & -25:28:43.72 & 0.91 & 14.35 & 5.42 &  31.69 &  5.00 & 15.28 & 0.453 \\
OB190726 & KB190853 & 17:51:43.04 & -31:08:50.89 & 1.09 & 14.69 & 5.71 & 281.97 &  9.06 & 12.66 & 0.814 \\
KB192073 & KB192073 & 17:49:53.08 & -29:35:17.30 & 0.94 & 14.45 & 5.43 &   4.77 &  6.41 & 11.17 & 0.152 \\
KB192528 & KB192528 & 17:39:03.05 & -33:20:43.91 & 1.81 & 13.13 &19.59 & 159.27 &  5.94 &  8.40 & 0.635 \\
KB192555 & KB192555 & 17:39:54.97 & -28:19:46.74 & 1.25 & 13.71 &10.04 &  33.44 &  5.31 &  6.89 & 0.334 \\
OB190551 & KB190519 & 17:59:28.75 & -28:50:25.80 & 1.49 & 12.61 &19.50 &   4.35 &  4.17 &  3.46 & 0.673 \\
OB190953 & KB191315 & 17:43:02.79 & -27:29:12.26 & 1.05 & 14.50 & 5.99 & 179.88 &  2.24 &  2.86 & 0.479 \\
KB191143 & KB191143 & 17:39:43.02 & -28:28:40.94 & 2.82 & 12.98 &30.10 & 489.43 &  2.88 &  2.34 & 0.077 \\
KB190527 & KB190527 & 17:27:54.38 & -26:55:30.22 & 0.99 & 14.74 & 5.05 &  71.03 &  1.45 &  1.65 & 0.998 \\
 \enddata
 \tablecomments{Event names are abbreviations for,
  e.g., KMT-2019-BLG-1820, OGLE-2019-BLG-1182, and 
  MOA-2019-BLG-006}
 \label{tab:fspl}
 \end{deluxetable}

%\begin{figure}
%\plotone{lc.eps}
%\caption{more stuff for KMT-2019-BLG-2073}
%\label{fig:}
%\end{figure}


\begin{thebibliography}{100}

%\bibitem[Afonso et al.(2001)]{eb20005} Afonso, C., Albert, J.N., Anderson, J. et al., 2001, \aap, 378, 1014

%aaa: added reference paper Agol, E. 2003, ApJ, 594, 449A
\bibitem[Agol(2003)]{agol03}Agol, E., 2003, \apj, 594, 449

\bibitem[Alard \& Lupton(1998)]{alard98} Alard, C. \& Lupton, R.H.,1998, \apj, 503, 325

%\bibitem[Albrow et al.(2000)]{mb9741}Albrow, M.\ Beaulieu, J.-P., Caldwell, J.A.R.,, et al.\ 2000, \apj, 534, 894

%\bibitem[Albrow et al.(2001)]{albrow01}Albrow, M.D., An, J., Beaulieu, J.-P., et al.\ 2000, \apjl, 556, 113

\bibitem[Albrow et al.(2009)]{albrow09}Albrow, M.\ D., Horne, K., Bramich, D.\ M., et al.\ 2009, \mnras, 397, 2099


%\bibitem[Alcock et al.(2001)]{alcock01} Alcock, C., Alssman, R.A., Alves, D.R., et al. 2001, Nature, 414, 617


%\bibitem[An \& Gould(2001)]{angould}An, J.H., \& Gould, A. 2001, \apj, 563, L111

%\bibitem[Alonso-Garc{\' \i}a et al.(2012)]{AlonsoGarcia2012}Alonso-Garc{\' \i}a, J, Mateo, M., Sen, B., et al. 2012, \aj, 143, 70

%\bibitem[Bachelet et al. (2012)]{bachelet12}Bachelet, E., Fouqu{\'e}, P., Han, C., et al.\ 2012, \aap, 547, 55

%\bibitem[Bachelet et al. (2012)]{mb10477}Bachelet, E., Shin, I.-G.., Han, C., et al.\ 2012, \apj, 754, 75


%\bibitem[Batista et al.(2011)]{mb09387} Batista, V., Gould, A., Dieters, S. et al. \aap, 529, 102

%\bibitem[Batista et al.(2014)]{mb11293b} Batista, V., Beaulieu, J.-P., Gould, A., et al. \apj, 780, 54

%\bibitem[Batista et al.(2015)]{ob05169bat} Batista, V., Beaulieu, J.-P., Bennett, D.P., et al. 2015, \apj, 808, 170

%\bibitem[Beaulieu et al.(2006)] {ob05390}Beaulieu, J.-P. Bennett, D.P., Fouqu\'e, P. et al. 2006, Nature, 439, 437

%\bibitem[Bennett \& Rhie(1996)]{bennett96}  Bennett, D.P. \& Rhie, S.-H. 1996, \apj, 472, 660

%\bibitem[Bennett et al.(2008)]{mb07192}  Bennett, D.P., Bond, I.A., Udalski, A., et al.\ 2008, \apj, 684, 663

%\bibitem[Bennett et al.(2010)]{ob06109b}Bennett, D.P., Rhie, S.H., Nikolaev, S. et al. 2010, \apj, 713, 837

%\bibitem[Bennett et al.(1999)]{mb9741x}  Bennett, D.P., Rhie, D.P., Becker, A.C., et al.\ 1999, Natur, 402 57

%\bibitem[Bennett et al.(2012)]{moabin1}  Bennett, D.P., Sumi, T., Bond, I.A., et al.\ 2012, \apj, 757, 119

%\bibitem[Bennett et al.(2015)]{ob05169ben} Bennett, D.P., Bhattacharya, A., Anderson, J., et al. 2015, \apj, 808, 169

%\bibitem[Bennett et al.(2016)]{ob07349} Bennett, D.P., Rhie, S.H., Udalski, A., et al. 2016, \aj, 152, 125

\bibitem[Bennett et al.(2020)]{ob05071c}Bennett, D. P., Bhattacharya, A., Beaulieu, J. P., et al. 2020, \aj, 159, 68

\bibitem[Bensby et al.(2013)]{bensby13} Bensby, T. Yee, J.C., Feltzing, S.\ et al.\ 2013, \aap, 549A, 147

\bibitem[Bessell \& Brett(1988)]{bb88} Bessell, M.S., \& Brett, J.M.\ 1988, \pasp, 100, 1134

%\bibitem[Bond et al.(2004)]{ob03235} Bond, I.A., Udalski, A., Jaroszy\'nski, M. et al.\ 2004, \apj, 606, L155

%\bibitem[Bond et al.(2017)]{ob161195a} Bond, I.A., Bennett, D.P., Sumi, T. et al.\ 2017, \mnras, 469, 2434

\bibitem[Bowler et al.(2015)]{bowler15} Bowler, B. P., Liu, M. C., Shkolnik, E. L., Tamura, M. 2015, \apjs, 216, 7

%\bibitem[Bozza et al.(2012)]{bozza12}Bozza, V., Dominik, M., Rattenbury, N.\ J., et al.\ 2012, \mnras, 424, 902

%\bibitem[Bramich(2008)]{bramich08} Bramich, D. M. 2008, \mnras, 386, L77.

%\bibitem[Calchi Novati et al.(2015)]{21event}  Calchi Novati, S., Gould, A., Udalski, A., et al., 2015, \apj, 804, 20

%\bibitem[Calchi Novati et al.(2015)]{170event} Calchi Novati, S., Gould, A., Yee, J.C., et al. 2015, \apj, 814, 92

%\bibitem[Cassan et al.(2012)]{cassan12} Cassan, A., Kubas, D., Beaulieu, J.-P., et al., 2012, Nature, 481, 167

%\bibitem[Chang-Refsdal(1979)]{cr1}Chang, K.\ \& Refsdal, S.\ 1979, Nature, 282, 561

%\bibitem[Chang \& Refsdal(1984)]{cr2}Chang, K.\ \& Refsdal, S.\ 1984, \aap, 130, 157

%\bibitem[Choi et al.(2013)]{choi13}Choi, J.-Y., Han, C., Udalski, A., et al.\ 2013, \apj, 768, 129

\bibitem[Chung et al.(2017)]{ob151482} Chung, S.-J., Zhu, W., Udalski, A., 2017, \apj, 838, 154

%\bibitem[Clanton \& Gaudi(2014a)]{clanton14}Clanton, C.D. \& Gaudi, B.S. 2014a, \apj, 791, 90

%\bibitem[Clanton \& Gaudi(2014b)]{clanton14b}Clanton, C.D. \& Gaudi, B.S. 2014b, \apj, 791, 91

%\bibitem[Clanton \& Gaudi(2016)]{clanton16}Clanton, C.D. \& Gaudi, B.S. 2016, \apj, 819, 125

%\bibitem[Claret(2000)]{claret00}Claret, A.\ 2000, \aap, 363,1081

%\bibitem[Dong et al.(2006)]{ob04343} Dong, S., DePoy, D.L., Gaudi, B.S., et al. 2006, \apj, 642, 842

%\bibitem[Dong et al.(2009)]{ob05071b} Dong, S., Gould, A., Udalski, A., et al. 2009, \apj, 695, 970

\bibitem[Dong et al.(2009)]{mb07400} Dong, S., Bond, I.A., Gould, A., et al. 2009, \apj, 695, 970

%\bibitem[Dominik(1999)]{dominik99} Dominik, M. 1999, \aap, 349, 108

%\bibitem[Dominik et al.(2007)]{signalmen} Dominik, M., Rattenbury, N.J., Allan, A., et al. 2007, \mnras, 380, 792.

%\bibitem[Evans et al.(2016)]{evans16} Evans, D.F., Southworth, J., Maxted, P.F.L., et al. 2016, \aap 589, 58.

%\bibitem[Fabrycky \& Tremaine(2007)]{fabrycky07} Fabrycky, D. \& Tremaine, S. 2007, \apj, 669, 1298

%\bibitem[Gaudi(1998)]{gaudi98} Gaudi, B.S.\ 1998, \apj, 506, 533

%\bibitem[Gaudi et al.(2008)]{ob06109}Gaudi, B.S., Bennett, D.P., Udalski, A. et al.\ 2008, Science, 319, 927

%\bibitem[Gaudi(2012)]{gaudi12} Gaudi, B.S.\ 2012, \araa, 50, 411

%\bibitem[Gaudi \& Gould(1997)]{gaudi97} Gaudi, B.S. \& Gould, A.\ 1997, \apj, 477, 152

%\bibitem[Gaudi \& Gould(1997)]{gaudi97} Gaudi, B.S. \& Gould, A.\ 1997, \apj, 486, 85

%\bibitem[Gaudi et al.(1998)]{gaudi-multi} Gaudi, B.S., Naber, R.M., \& Sackett, P. 1998, \apj, 502, L33

%\bibitem[Gaudi \& Sackett(2000)]{gaudi00} Gaudi, B.S. \& Sackett, P. 2000, \apj, 528, 56

%\bibitem[Gaudi et al.(2002)]{gaudi02} Gaudi, B.S., Albrow, M.D., An, J.\ 2002, \apj, 566, 463

\bibitem[Gonzalez et al.(2012)]{gonzalez12} Gonzalez, O.~A., Rejkuba, M., Zoccali, M., et al.\ 2012, \aap, 543, A13

%\bibitem[Gould(1992)]{gould92} Gould, A. 1992, \apj, 392, 442

%\bibitem[Gould(1994a)]{gould94a} Gould, A. 1994a, \apjl, 421, L71

%\bibitem[Gould(1994b)]{gould94b} Gould, A. 1994b, \apjl, 421, L75

%\bibitem[Gould(1995)]{gould95} Gould, A. 1995, \apjl, 441, L21

\bibitem[Gould(1996)]{gould96} Gould, A. 1996, \apj, 470, 201

%\bibitem[Gould(1997)]{gould97} Gould, A. 1997, The Hollywood Strategy for Microlensing Detection of Planets, in Variables Stars and the Astrophysical Returns of the Microlensing Surveys. Eds. R. Ferlet, J.-P. Maillard and B. Raban. Gif-sur-Yvette, France : Editions Frontieres, p.125

%\bibitem[Gould(2000)]{gould00} Gould, A. 2000, \apj, 542, 785

%\bibitem[Gould(2004)]{gould04} Gould, A. 2004, \apjl, 606, 319

\bibitem[Gould(2008)]{gould08} Gould, A.\ 2008, \apj, 681, 1593

\bibitem[Gould (2016)]{gould16} Gould, A. 2016, JKAS, 49, 123

\bibitem[Gould(2019)]{gould19} Gould, A. 2019, JKAS, 52, 121

\bibitem[Gould et al. (2021)]{gould21} Gould, A., Zang W., Mao, S., Dong, S. 2021, RAA, in press

%\bibitem[Gould \& Andronov(1999)]{gouldandronov99} Gould, A.\ \& Andronov, N. 1999, \apj, 516, 236

%\bibitem[Gould et al.(2004)]{gba04} Gould, A., Bennet, D.P., \& Alves, D.R., \apj, 614, 404


%\bibitem[Gould \& Gaucherel(1997)]{gg97} Gould, A. \& Gaucherel. 1997, \apj, 477, 580

%\bibitem[Gould \& Horne(2013)]{gouldhorne} Gould, A. \& Horne, K. 2013, \apjl, 779, L28

%\bibitem[Gould \& Loeb(1992)]{gouldloeb} Gould, A. \& Loeb, A. 1992, \apj, 396, 104

\bibitem[Gould \& Yee(2012)]{gould12} Gould, A. \& Yee, J.C. 2012, \apj, 755, L17

%\bibitem[Gould et al.(2006)]{ob05169} Gould, A., Udalski, A., An, D.\ et al.\ 2006, \apj, 644, L37

%\bibitem[Gould et al.(2009)]{ob07224} Gould, A., Udalski, A., Monard, B.\ et al.\ 2013, \apj, 698, L147

%\bibitem[Gould et al.(2014)]{ob130341} Gould, A., Udalski, A., Shin, I.-G.\ et al.\ 2014, Science, 345, 46


%\bibitem[Gould et al.(2013)]{prop2013} Gould, A., Carey, S., \& Yee, J. 2013, 2013spitz.prop.10036

%\bibitem[Gould et al.(2014)]{prop2014} Gould, A., Carey, S., \& Yee, J. 2014, 2014spitz.prop.11006

%\bibitem[Gould et al.(2015a)]{prop2015a} Gould, A., Yee, J., \& Carey, S., 2015a, 2015spitz.prop.12013

%\bibitem[Gould et al.(2015b)]{prop2015b} Gould, A., Yee, J., \& Carey, S., 2015b, 2015spitz.prop.12015

%\bibitem[Gould et al.(2016)]{prop2016} Gould, A., Yee, J., \& Carey, S., 2016, 2015spitz.prop.13005

%\bibitem[Gould et al.(2010)]{gould10} Gould, A., Dong, S., Gaudi, B.S.\ et al.\ 2010, \apj, 720, 1073

%\bibitem[Gould et al.(2013)]{gould13} Gould, A., Shin, I.-G., Han, C. et al.\ 2013, \apj, 768, 126

%\bibitem[Grether \& Lineweaver(2006)]{grether06} Grether, D., \& Lineweaver, C.H., 2006, \apj, 640, 1051

%\bibitem[Griest \& Safizadeh(1998)]{griest98} Griest, K.\ \& Safizadeh, N.\ 1998, \apj, 500, 37

\bibitem[Groenewegen(2004)]{groenewegen04} Groenewegen, M.A.T., 2004, \mnras, 353, 903

%\bibitem[Han \& Gould(1995)]{han95} Han, C. \& Gould, A.\ 1995, \apj, 447, 53

%\bibitem[Han \& Gould(2003)]{han03} Han, C. \& Gould, A.\ 2003, \apj, 592, 172

%\bibitem[Han(2006)]{han06} Han, C.  2006, \apj, 638, 1080

%\bibitem[Han et al.(2001)]{planet-factor} Han, C., Chang, H.-Y., An, J.H., \& Chang, K.  2001, \mnras, 328, 986

%\bibitem[Han et al.(2013)]{ob120026}Han, C., Udalski, A., Lee, Choi, J.-Y., et al. 2013, \apj, 778, 38

%\bibitem[Han et al.(2016)]{ob150768}Han, C., Udalski, A., Lee, C.-U., et al. 2016, \apj, 827, 11

%\bibitem[Han et al.(2017)]{ob160613} Han, C., Udalski, A., Gould, A.\ 2017, \aj, 154, 223

\bibitem[Han et al.(2020)]{mb19256} Han, C., Lee, C.-U., et al.\ 2020, \aj, 159, 134

%\bibitem[Henderson et al.(2016)]{henderson16} Henderson, C.B., Poleski, R., Penny, M. et al. 2016 \pasp, 128, 124401

\bibitem[Hirao et al.(2020)]{ob170406} Hirao, Y., Bennett, D.P., Ryu, Y.-H., et al., 2020, in press

%\bibitem[Hodgkin et al.(2009)]{Hodgkin.2009.A} Hodgkin, S.~T., Irwin, M.~J., Hewett, P.~C., \& Warren, S.~J.\ 2009, \mnras, 394, 675

%\bibitem[Hwang et al.(2018)]{ob170173} Hwang, K.-H., Udalski, A., Shvartzvald, Y. et al. 2018, \aj, 155, 20 %, arXiv:1709.08476

%\bibitem[Hwang et al.(2018)]{kb160212} Hwang, K.-H., Kim, H.-W., Kim, D.-J., et al. 2018, JKAS, 155, 20 

%\bibitem[Hwang et al.(2018)]{kb161107} Hwang, K.-H., Ryu, Y.-H., Kim, H.-W., et al. 2018, AAS submitted

%\bibitem[Hwang et al.(2019)]{ob151459} Hwang, K.-H., Udalski, A., Bond, I.A., et al. 2018, \aj, 155,259

%\bibitem[Irwin et al.(2004)]{Irwin.2004.A} Irwin, M.~J., Lewis, J., Hodgkin, S., et al.\ 2004, \procspie, 5493, 411

%\bibitem[Jung et al.(2014)]{jung14}Jung, Y.\ K., Park, H., Han, C., et al. 2014 \apj, 786, 85

%\bibitem[Janczak et al.(2010)]{mb08310}Janczak, J., Fukui, A., Dong, S., et al.\ \apj, 711, 731

\bibitem[Johnson et al. (2020)]{johnson20}Johnson, S.A., Penny, M., Gaudi, B.S., et al. 2020, \aj, 160, 123

%\bibitem[Jung et al.(2015)]{jung15}Jung, Y.\ K., Udalski, A., Sumi, T., et al.\ 2015 \apj, 798, 123

%\bibitem[Jung et al.(2013)]{jung13}Jung, Y.\ K., Han, C., Gould, A., \& Maoz, D. 2013 \apjl, 768, L7

%\bibitem[Jung et al.(2019)]{kb170165}Jung, Y.\ K., Gould, A., \& Zang, W. et al. 2019, \aj, 157, 72

%\bibitem[Kayser et al.(1986)]{kayser86} Kayser, R., Refsdal, S., \& Stabell, R. 1986, \aap, 166, 36

%\bibitem[Kennedy \& Kenyon(2008)]{kennedy08} Kennedy, G.M. \& Kenyon, S.J. 2008, \apj, 673, 502

\bibitem[Kervella et al.(2004)]{kervella04} Kervella, P., Th{\'e}venin, F., Di Folco, E., \& S{\'e}gransan, D.\ 2004, \aap, 426, 297

\bibitem[Kim et al.(2016)]{kmtnet} Kim, S.-L., Lee, C.-U., Park, B.-G., et al.  2016, JKAS, 49, 37

\bibitem[Kim et al.(2018a)]{eventfinder} Kim, D.-J., Kim,  H.-W., Hwang, K.-H., et al., 2018a, \aj, 155, 76

\bibitem[Kim et al.(2018b)]{alertfinder} Kim, D.-J., Hwang, K.-H., Shvartzvald, et al. 2018b, arXiv:1806.07545

%\bibitem[Kim et al.(2018b)]{2016k2} Kim,  H.-W., Hwang, K.-H., Kim, D.-J., et al., 2018b, \aj, 155, 186

%\bibitem[Kim et al.(2018c)]{2016eventfinder} Kim,  H.-W., Hwang, K.-H., Kim, D.-J., et al., 2018c, AAS submitted, arXiv:1804.03352

%\bibitem[Kubas et al.(2012)]{mb07192k}Kubas, D., Beaulieu, J.-P., Bennett, D.P., et al. 2012, \aap, 540A, 78

%aaa: added reference paper Maeder, A. 1973, A&A, 26, 215M
\bibitem[Maeder(1973)]{maeder73}Maeder, A., 1973, \aap, 26, 215

%\bibitem[Marcy \& Butler(2000)]{marcy00} Marcy, G.W. \& Butler, R.P. 2000, \pasp, 112,137

%\bibitem[Mao \& Paczy\'nski(1991)]{mao91} Mao, S.\ \& Paczy\'nski, B.\ 1991, \apj, 374, 37

\bibitem[Minniti et al.(2017)]{vvvcat}Minniti, D., Lucas, P., VVV Team, 2017, yCAT 2348, 0

%\bibitem[Miyake et al.(2011)]{mb09319} Miyake, N., Sumi, T., Dong, S. et al. \apj, 728, 120

%\bibitem[Mr\'oz et al.(2017)]{ob160596}Mr\'oz, P., Han, C., Udalski, A. et al.\ 2017, \aj, 153, 143

\bibitem[Mr\'oz et al.(2017)]{mroz17}Mr\'oz, P., Udalski, A., Skowron, J., et al.\ 2017, Nature, 548, 183

\bibitem[Mr\'oz et al.(2018)]{ob161540}Mr\'oz, P., Ryu, Y.-H., Skowron, J., et al.\ 2018, \aj, 155, 121

\bibitem[Mr\'oz et al.(2019)]{ob121323}Mr\'oz, P., Udalski, A., Bennett, D.P.., et al.\ 2019, \aap, 622, A201

\bibitem[Mr\'oz et al.(2020)]{ob190551}Mr\'oz, P., Poleski, R., Han, C., et al.\ 2020, \aj, 159, 262


%\bibitem[Muraki et al.(2011)]{mb09266} Muraki, Y., Han, C., Bennett, D.P., et al.\ 2011, \apj, 741, 22

\bibitem[Nataf et al.(2013)]{nataf13} Nataf, D.M., Gould, A., Fouqu\'e, P. et al. 2013, \apj, 769, 88

%\bibitem[Pascucci et al.(2018)]{kepq}Pascucci, I., Mulders, G.D., Gould, A., \& Fernandes, R. 2018, \apjl, 856, L28

\bibitem[Paczy\'nski(1986)]{pac86} Paczy\'nski, B.\ 1986, \apj, 304, 1

%\bibitem[P\'al(2012)]{Pal2012} P\'al, A.\ 2012, \mnras, 421, 1825

%\bibitem[Park et al.(2013)]{park13}Park, H., Udalski, A., Han, C., et al.\ 2013, \apj, 778, 134

%\bibitem[Park et al.(2015)]{park15}Park, H., Udalski, A., Han, C., et al.\ 2015, \apj, 805, 117

\bibitem[Pejcha \& Heyrovsk\'y(2009)]{pejcha09} Pejcha, O., \& Heyrovsk\'y, D.\ 2009, \apj, 690, 1772

%\bibitem[Penny et al.(2016)]{penny16} Penny, M.T., Henderson, C.B., \& Clanton, C.\ 2016, \apj, 830, 150

%\bibitem[Poleski et al.(2014)]{ob08092} Poleski, R., Skowron, J., Udalski, A.,  et al.\ 2014, \apj, 755, 42

%\bibitem[Poleski et al.(2017)]{mb12006} Poleski, R., Udalski, A.,  Bond, I.A., et al.\ 2017, \aap, 604A, 103

%\bibitem[Poleski et al.(2014b)]{ob120406} Poleski, R., Udalski, A., Dong, S.\ et al.\ 2014b, \apj, 782, 47

%\bibitem[Poleski et al.(2014b)]{mb12006} Poleski, R., Udalski, A., Bond, I.A..\ et al.\ 2017, arXiv:1704.01121

\bibitem[Poleski et al.(2018)]{poleski18} Poleski, R., Gaudi, B.S., Udalski, A. et al. 2018, \aj, 156, 104

\bibitem[Poleski et al.(2020)]{poleski20} Poleski, R., et al., 2020, in prep

%\bibitem[Ranc et al.(2018)]{ob151670} Ranc, C., Bennett, D.P., Hirao, Y., et al.\ 2018 arXiv:1810.00014

%\bibitem[Refsdal(1966)]{refsdal66} Refsdal, S. 1966, \mnras, 134, 315

%\bibitem[Rhie et al.(2000)]{rhie00} Rhie, S.H., Bennett, D.P., Becker, A.C. et al. 2000, ApJ, 533, 378

%aaa: added reference paper Riffeser A., Fliri, J., Seitz, S. & Bender, R. 2006, ApJS, 163, 225S
\bibitem[Riffeser et al.(2006)]{riffeser06} Riffeser, A., Fliri, J., Seitz, S., \& Bender, R. 2006, \apjs, 163, 225

%\bibitem[Ryu et al.(2017)]{ob160693} Ryu, Y.-H., Udalski, A., Yee, J.C. et al. 2017, \aj, 154, 247 %arXiv:1707.12222

\bibitem[Ryu et al.(2019a)]{kb181292} Ryu, Y.-H., Navarro, M.G., Gould, A. et al. 2019, \aj, 159, 58

\bibitem[Ryu et al.(2021)]{kb172820} Ryu, Y.-H., Mr\'oz, P., Gould, A. et al. 2021, \aj, 161, 126

%\bibitem[Santos et al.(2004)]{santos04} Santos, N.~C., Israelian, G., \& Mayor, M.\ 2004, \aap, 415, 1153

%\bibitem[Santerne et al.(2016)]{santerne16} Santerne, A., Beaulieu, J.-P., Rojas Ayala, B., et al. 2016, \aap, in press  arXiv:1610.04446


\bibitem[Schechter et al.(1993)]{dophot} Schechter, P.L., Mateo, M., \& Saha, A. 1993, \pasp, 105, 1342

%\bibitem[Schmidt(1968)]{schmidt68} Schmidt, M., 1968 \apj, 151, 393

%\bibitem[Schneider \& Weiss(1988)]{schneider88}Schneider, P., \& Weiss, A. 1988, \apj, 330, 1

\bibitem[Schlafly et al.(2018)]{decam}Schlafly, E.F., Green, G.M., Lang, D. et al. 2018, \apjs, 234, 39

%\bibitem[Shen et al.(2018)]{ob140962}Shen, Y., Yee, J.C., Udalski, A., et al.\ 2018, arXiv:1805.09350

%\bibitem[Shin et al.(2012a)]{ob110417}Shin, I.-G., Han, C., Choi, J.-Y., et al.\ 2012a, \apj, 755, 91

%\bibitem[Shin et al.(2012b)]{shin12b}Shin, I.-G., Han, C., Gould, A., et al.\ 2012b, \apj, 760, 116


%\bibitem[Shin et al.(2015)]{shin-3L1S}Shin, I.-G., Han, C., Choi, J.-Y.., et al.\ 2015, \apj, 802, 108


%\bibitem[Shin et al.(2016)]{ob150954} Shin, I.-G., Ryu, Y.H, Udalski, A.\ et al. 2016, JKAS, 49, 73

%\bibitem[Shvartzvald et al.(2014)]{mb11322} Shvartzvald, Y., Maoz, D., Kaspi, S.\ et al.\ 2014, \mnras, 439, 604

%\bibitem[Shvartzvald et al.(2016)]{shvartzvald16} Shvartzvald, Y., Maoz, D., Udalski, A.\ et al.\ 2016, \mnras, 457, 4089

%\bibitem[Shvartzvald et al.(2017)]{Shvartzvald.2017.A} Shvartzvald, Y.,Bryden, G., Gould, A., et al.\ 2017, \aj, 153, 61


%\bibitem[Shvartzvald et al.(2015)]{ob151285} Shvartzvald, Y., Udalski, A., Gould, A.\ et al.\ 2015, \apj, 814, 111

%\bibitem[Shvartzvald et al.(2017a)]{ukirt17} Shvartzvald, Y., Bryden, G.,Gould, A.\ et al.\ 2017a, \aj, 135, 61

%\bibitem[Shvartzvald et al.(2017)]{ob161195b} Shvartzvald, Y., Yee, J.C., Calchi Novati, S.\ et al.\ 2017, \apjl, 840, L3

%\bibitem[Skottfelt et al.(2015)]{skottfelt15} Skottfelt, J., Bramich, D. M., Hundertmark, M., et al., 2015, \aap, 574, 54



%\bibitem[Skowron et al.(2011)]{ob09020}Skowron, J., Udalski, A., Gould, A et al.\ 2011, \apj, 738, 87

%\bibitem[Skowron et al.(2011)]{mb11028} Skowron, J., Udalski, A., Poleski, R. et al.\ 2016, \apj, 820, 4

%\bibitem[Skowron et al.(2016)]{skowron16} Skowron, J., Udalski, A., Koz{\l}owski, S. et al.\ 2016, Acta Astron., 66, 1

%\bibitem[Skowron et al.(2018)]{ob170373} Skowron, J., Ryu, Y.-H., Hwang, K.-H., et al.\ 2016, Acta Astron., 68, 43

%\bibitem[Smith et al.(2003)Smith, Mao \& Paczy\'nski]{smp03} Smith, M., Mao, S., \& Paczy\'nski, B., 2003, \mnras, 339, 925

%\bibitem[Song et al.(2014)]{song14} Song, Y.-Y., Mao, S., \& An, J.H. 2014, \mnras, 437, 4006

%\bibitem[Spergel et al.(2013)]{spergel13} Spergel, D.N., Gehrels, N., Breckinridge, J., et al. 2013, arXiv:1305.5422

%\bibitem[Street et al.(2013)]{street13}Street, R.\ A., Choi, J.-Y., Tsapras, Y., et al.\ 2013, \apj, 763, 67

%\bibitem[Street et al.(2016)]{ob150966} Street, R., Udalski, A., Calchi Novati, S.\ et al.\ 2016, \apj, 829, 93.

%\bibitem[Sumi et al.(2010)]{ob07368} Sumi, T., Bennett, D.P., Bond, I.A., et al.\ 2010, \apj, 710, 1641

\bibitem[Sumi et al.(2011)]{sumi11} Sumi, T., Kamiya, K., Bennett, D.~P., et al.\ 2011, Nature, 473, 349 

%\bibitem[Sumi et al.(2016)]{mb13605} Sumi, T., Udalski, A., Bennett, D.P., et al.\ 2016 \apj, in press arXiv:1512.00134

%\bibitem[Suzuki et al.(2016)]{suzuki16} Suzuki, D., Bennett, D.P., Sumi, T., et al. 2016, \apj, 833, 145

%\bibitem[Suzuki et al.(2018)]{ob141722} Suzuki, D., Bennett, D.P., Udalski, A., et al. 2018, \aj, 155, 263

\bibitem[Szyma\'nski et al.(2011)]{oiiicat}Szyma\'nski, M.K., Udalski, A., Soszy\'nski, I., et al. 2011, Acta Astron., 61, 83

%\bibitem[Thompson(2013)]{thompson13} Thompson, T.A. 2013, \mnras, 431, 63

\bibitem[Tomaney \& Crotts(1996)]{tomaney96} Tomaney, A.B. \& Crotts, A.P.S. 1996, \au, 112, 2872

%\bibitem[Tsapras et al.(2014)]{ob120406b}Tsapras, Y., Choi, J.-Y., Street, R.-A., et al. 2014, \apj, 782, 48

%\bibitem[Udalski(2003)]{ews2} Udalski, A. 2003, Acta Astron., 53, 291

%\bibitem[Udalski et al.(1994)]{ews1} Udalski, A.,Szymanski, M., Kaluzny, J., Kubiak, M., Mateo, M.,  Krzeminski, W., \& Paczy\'nski, B. 1994, Acta Astron., 44, 227

%\bibitem[Udalski et al.(2005)]{ob05071} Udalski, A., Jaroszy\'nski, M., Paczy\'nski, B, et al. 2005, \apj, 628, L109.

%\bibitem[Udalski et al.(2015)]{ob140124} Udalski, A., Yee, J.C., Gould, A., et al. 2015, \apj, 799, 237

%\bibitem[Udalski et al.(2015)]{ogle-iv} Udalski, A., Szymanski, M.K., Szymanski, G., et al. 2015, Acta Astron., 65, 1

%\bibitem[Udalski et al.(2018)]{ob171434} Udalski, A., Ryu, Y.-H., Sajadian, S., et al.\ 2018, Acta Astron., 68, 1

%\bibitem[Fischer \& Valenti(2005)]{fischer05} Fischer, D.~A., \& Valenti, J.\ 2005, \apj, 622, 1102


%\bibitem[Wambsganss(1997)]{wambs97}Wambsganss, J. 1997, \mnras, 284, 172

%\bibitem[Wo\'zniak(2000)]{wozniak2000} Wo\'zniak, P.~R. 2000, Acta Astron., 50, 421

%\bibitem[Wyrzykowski et al.(2015)]{wyr15}Wyrzykowski, \L., Rynkiewicz, A.E., Skowron, J., et al. 2015, \apjs, 216, 12

%\bibitem[Udalski et al.(2015b)]{ogleiv} Udalski, A., Szyma\'nski, M.K. \& Szyma\'nski, G. 2015b, Acta Astronom., 65, 1

%\bibitem[Wu(2018)]{wu18} Wu, Y. 2018, arXiv:1806.04693

%\bibitem[Yee et al.(2012)]{mb11293} Yee, J.C., Shvartzvald, Y., Gal-Yam, A.\ et al.\ 2012, \apj, 755, 102

%\bibitem[Yee et al.(2013)]{mb10311} Yee, J.C., Hung, L.-W., Bond, I.A.\ et al.\ 2013, \apj, 769, 77

%\bibitem[Yee et al.(2015)]{yee15} Yee, J.C., Gould, A., Beichman, C., 2015, \apj, 810, 155

%\bibitem[Yee et al.(2016)]{yee16} Yee, J.C., Johnson, J.A., Skowron, J., et al. 2016, \apj, 821, 121

\bibitem[Yoo et al.(2004)]{ob03262} Yoo, J., DePoy, D.L., Gal-Yam, A.\ et al.\ 2004, \apj, 603, 139

%\bibitem[Zang et al.(2018)]{cfht-phot} Zang, W., Penny, M.T., Zhu, W., et al. 2018, \pasp, 130, 104401

%\bibitem[Zhu et al.(2014)]{Zhu:2014} Zhu, W., Penny, M., Mao, S., Gould, A., \& Gendron, R.\ 2014, \apj, 788, 73

%\bibitem[Zhu et al.(2014)]{zhu-2planet} Zhu, W., Gould, A., Penny, M., et al.\ 2014, \apj, 794, 53

%\bibitem[Zhu et al.(2015)]{zhu15} Zhu, W., Gould, A., XXX, et al. 2015, \apj, 814, 129

%\bibitem[Zhu et al.(2015)]{ob141050} Zhu, W., Udalski, A., Gould, A., et al. 2015, \apj, 805, 8

%\bibitem[Zhu et al.(2016)]{ob150763} Zhu, W., Calchi Novati, S., Gould, A., et al. 2016, \apj, 825, 60

%\bibitem[Zhu et al.(2017)]{zhu17} Zhu, W., Udalski, A., Calchi Novati, S.,  et al. 2017, \aj, 154, 210 %arXiv:1701.05191

%\bibitem[Zhu et al.(2018a)]{zhu18a} Zhu, W., Petrovich, C.,  Wu, Y, \& Ha, Y.-C., 2018b, \apj, 860, 101

%\bibitem[Zhu et al.(2018b)]{zhu18b} Zhu, W. \& Wu, Y, 2018b, arXiv:18050.26660

\end{thebibliography}
\end{document}